\documentclass[12pt,letterpaper]{article}
\pdfoutput=1
\usepackage[includeheadfoot,
            marginratio={1:1,2:3}, 
            width=412pt, 
            height=688pt,]{geometry}

\usepackage{amsmath}
\usepackage{amsfonts}
\usepackage{amssymb}
\usepackage{graphicx}
\usepackage{cite}
\usepackage{ulem}
\usepackage{dsfont}
\usepackage{multirow}

\newcommand{\eq}[1]{\begin{equation}
                     \begin{split} #1 \end{split}
                     \end{equation}}

\newcommand{\ov}{\overline}
\newcommand{\op}{\hspace{1pt}}
\newcommand{\lab}{\mathsf }

\allowdisplaybreaks[2]
\numberwithin{equation}{section}

\begin{document}

\normalem
\vspace*{-1.5cm}
\begin{flushright}
  {\small
  MPP-2016-177 \\
  LMU-ASC 36/16
  }
\end{flushright}

\vspace{1.5cm}

\begin{center}
{\LARGE
Partial SUSY Breaking for Asymmetric Gepner  \\[0.2cm]
Models and Non-geometric Flux Vacua\\[0.3cm]
}
\vspace{0.4cm}

\end{center}

\vspace{0.35cm}
\begin{center}
  Ralph Blumenhagen$^1$, Michael Fuchs$^1$, Erik Plauschinn$^2$
\end{center}

\vspace{0.1cm}
\begin{center} 
\emph{
$^{1}$ Max-Planck-Institut f\"ur Physik (Werner-Heisenberg-Institut), \\ 
F\"ohringer Ring 6,  80805 M\"unchen, Germany 
} \\[1em] 

\emph{
$^{2}$ Arnold Sommerfeld Center for Theoretical Physics,\\ 
LMU, Theresienstr.~37, 80333 M\"unchen, Germany
}
\end{center} 

\vspace{1cm}

%%%%%%%%%%%%%%%%%%%%%%%%%%%%%%%%%%%%%%%%%%%%%%%
%%%%%%%%%%%%%%%%%%%%%%%%%%%%%%%%%%%%%%%%%%%%%%%
%%%%%%%%%%%%%%%%%%%%%%%%%%%%%%%%%%%%%%%%%%%%%%%
%%%%%%%%%%%%%%%%%%%%%%%%%%%%%%%%%%%%%%%%%%%%%%%
%%%%%%%%%%%%%%%%%%%%%%%%%%%%%%%%%%%%%%%%%%%%%%%
%%%%%%%%%%%%%%%%%%%%%%%%%%%%%%%%%%%%%%%%%%%%%%%
%%%%%%%%%%%%%%%%%%%%%%%%%%%%%%%%%%%%%%%%%%%%%%%
%%%%%%%%%%%%%%%%%%%%%%%%%%%%%%%%%%%%%%%%%%%%%%%

\begin{abstract}
\noindent
Using the method of simple current extensions, asymmetric Gepner models of Type
IIB with ${\cal N}=1$ space-time supersymmetry are constructed.
The combinatorics of the massless vector fields suggests  
that these classical Minkowski string vacua  provide
fully backreacted solutions corresponding to ${\cal N}=1$ minima
of  ${\cal N}=2$ gauged supergravity. 
The latter contain abelian gaugings
along the axionic isometries  in the hypermultiplet moduli space, and
can be considered as  Type IIB flux
compactifications on Calabi-Yau  manifolds equipped with \mbox{(non-)}geometric
fluxes.
For a particular class of  asymmetric Gepner models,  we are able to explicitly specify the
underlying CICYs and to check necessary conditions for 
a GSUGRA interpretation. If this conjecture is correct, there exists a
large  class of exactly solvable non-geometric flux compactifications 
on CY threefolds.
\end{abstract}

%%%%%%%%%%%%%%%%%%%%%%%%%%%%%%%%%%%%%%%%%%%%%%%
%%%%%%%%%%%%%%%%%%%%%%%%%%%%%%%%%%%%%%%%%%%%%%%

\clearpage
\tableofcontents

\section{Introduction}

Classical vacua of string theory are described by two-dimensional
conformal field theories (CFTs). Space-time supersymmetry is implemented
by the spectral flow of the $N=2$ superconformal field theory (SCFT) on
the world-sheet, and all perturbative and non-perturbative $\alpha'$-corrections are incorporated 
in the CFT, whereas string-loop and non-perturbative string corrections
in the form of D-brane instantons have to be considered in addition.
Moreover, it is believed that CFT backgrounds  incorporate Neveu-Schwarz--Neveu-Schwarz (NS-NS)
fluxes but no Ramond--Ramond (R-R) fluxes.

These basic paradigms of string theory define the ideal approach
for the construction of (classical) solutions.
However, for studying string vacua one usually starts with
an effective supergravity theory at lowest order (in $\alpha'$)
and looks for solutions, either arguing or hoping that they
 extend to the full string equations of motion. 
In fact, the set of such leading-order solutions and the set
of exactly solvable SCFTs are fairly disjoint.
Only a few classes are known where agreement has been achieved,
mostly by indirect arguments.

The best controlled class is build upon toroidal compactifications
and orbifolds thereof, which are related to (quotients of)  free conformal field theories.
Another class is given by Gepner models
\cite{Gepner:1987qi,Gepner:1987vz}. 
This is a construction of $N=2$ SCFTs that are argued to provide exact
solutions to a string propagating on a Calabi-Yau (CY) manifold
at a very specific point in its moduli space. This point is deep
inside the K\"ahler moduli space, where sizes are of the order
of the string-scale $\sqrt{\alpha'}$. In view of the fact that
the Ricci-flat metric on the CY and therefore the corresponding
non-linear sigma model cannot even be written down, this
is still quite a remarkable result. This correspondence has been
argued for by comparison of massless spectra, the chiral ring  and
discrete symmetries. 

Using the method of simple current
extensions of partition functions
\cite{Schellekens:1989am,Schellekens:1989dq}, 
this construction was generalized to $(0,2)$
heterotic SCFT models in \cite{Blumenhagen:1995tt}, 
which were argued to correspond
to monad bundles with structure group $SU(n)$ on complete intersection
Calabi-Yau (CICY) manifolds
\cite{Blumenhagen:1995ew,Blumenhagen:1996vu}. 
However, for most of the solutions
of the leading-order SUGRA equations of motion the exact CFT is not
known. Moreover, there exist large classes of exactly known
SCFTs, which so far did not found an interpretation in terms
supergravity solutions. Among these are e.g.~the in general asymmetric
simple current extensions of Gepner models reported in
\cite{Schellekens:1989wx} (see also \cite{GatoRivera:2010gv,GatoRivera:2010xn}).

A large class of vacua, the so-called string landscape,   arises
in flux compactifications of string theory. Starting with for instance a
Calabi-Yau manifold, some of the moduli can be stabilized
by such fluxes, which is important for applications to string
phenomenology  and string cosmology.
Turning on fluxes leads to an ${\cal N}=2$  gauged supergravity (GSUGRA) theory whose vacuum structure can be quite involved. In particular,
the most generic gaugings include so-called non-geometric
fluxes. Such gaugings in ${\cal N}=2$ GSUGRA can be described
in the framework of $SU(3)\times SU(3)$ structures 
\cite{Grana:2005sn,Grana:2005ny,Benmachiche:2006df,Grana:2006hr,Cassani:2007pq}.
These can also be  described in  double field theory 
(for reviews see \cite{Aldazabal:2013sca,Berman:2013eva,Hohm:2013bwa})
and their
understanding has been under investigation during recent years \cite{Blumenhagen:2015lta}.
For instance, starting with a Type II compactification on  a
Calabi-Yau threefold, turning on generic NS-NS and R-R fluxes leads to an ${\cal
  N}=2$ gauged supergravity theory, where abelian isometries
along the axionic directions in the hypermultiplet moduli space are gauged
\cite{Louis:2002ny,Dall'Agata:2004nw,D'Auria:2004tr,D'Auria:2004wd,Grana:2005ny,D'Auria:2007ay}.

For a long time it was not clear whether this GSUGRA theory admits  
${\cal N}=1$ minima, i.e.~whether partial supersymmetry breaking
from ${\cal N}=2$ to ${\cal N}=1$ is possible. 
The first example was constructed in \cite{Ferrara:1995gu},
and in a more general context this issue was resolved
in the series of papers
\cite{Cassani:2009na,Louis:2009xd,Louis:2010ui,Hansen:2013dda}.  The main result is that for
partially-broken Minkowski vacua one needs at least two gauged
isometries through both electric and magnetic fluxes, with one
of them being non-geometric. Of course this is just a solution
to the GSUGRA equations of motion and it is not clear 
whether they uplift to genuine SCFTs. In fact, from a string theory
perspective one expects that 
turning on general types of fluxes leads to a strong backreaction\cite{Blumenhagen:2015kja}.
In particular, without having a dilute flux limit available for non-geometric fluxes, it is not  clear
whether the ansatz of a CY with constant fluxes can capture
the true vacuum structure.

In this paper we consider a subclass of the asymmetric simple current
extensions of Gepner models and suggest   
that these can be identified
with the fully backreacted solution of partially SUSY breaking
Minkowski vacua of  ${\cal N}=2$ GSUGRA.
We cannot prove this conjecture, but we 
collect some evidence for it.
In particular, for a set of such asymmetric CFTs (ACFTs) we 
give an explicit proposal
for the underlying CICY manifold of the ${\cal N}=2$ GSUGRA.
As in the $(0,2)$ setting \cite{Blumenhagen:1995ew,Blumenhagen:1996vu}, 
the guiding principle is the combinatorics of
the massless states in the asymmetric Gepner model, which reveals information about the weight
of the coordinates of an underlying weighted projective space.
If our conjecture is correct, it has interesting consequences:
\begin{itemize}
\item{Partial supersymmetry breaking is possible in string theory even
    beyond leading order in $\alpha'$.}
\item{Minima of an ${\cal N}=2$ GSUGRA theory 
can
correspond
    to classical minima of string theory.}
\item{Non-geometric fluxes are 
  part of the string degrees of
    freedom and correspond to ACFTs. This correspondence is also obtained for asymmetric orbifolds of tori\cite{Dabholkar:2002sy,Flournoy:2005xe,Condeescu:2012sp,Condeescu:2013yma}.}
\end{itemize}

Such asymmetric Gepner models \cite{Schellekens:1989wx} have also been
considered more recently  in the two papers \cite{Israel:2013wwa,Israel:2015efa},
so that our approach should be considered as an extension of their work.
In the present paper we go beyond them in two aspects.
First, we allow more general simple currents and second we do suggest that the ACFT
constructions are related to GSUGRA minima with partial SUSY breaking.

This paper is organized as follows: in section~\ref{sec_gep} we briefly review 
the construction of Gepner models and their partial breaking to $\mathcal N=1$ supersymmetry.
In section~\ref{sec_sugra} we summarize partial SUSY breaking from a 
supergravity point of view, and derive bounds on the spectrum after the breaking. 
In section~\ref{sec_examples} we discuss explicit examples for
our proposed ACFT -- GSUGRA correspondence, and section~\ref{sec_concl}
contains our conclusions.

%%%%%%%%%%%%%%%%%%%%%%%%%%%%%%%%%%%%%%%%%%%%%%%
%%%%%%%%%%%%%%%%%%%%%%%%%%%%%%%%%%%%%%%%%%%%%%%
%%%%%%%%%%%%%%%%%%%%%%%%%%%%%%%%%%%%%%%%%%%%%%%
%%%%%%%%%%%%%%%%%%%%%%%%%%%%%%%%%%%%%%%%%%%%%%%
%%%%%%%%%%%%%%%%%%%%%%%%%%%%%%%%%%%%%%%%%%%%%%%
%%%%%%%%%%%%%%%%%%%%%%%%%%%%%%%%%%%%%%%%%%%%%%%
%%%%%%%%%%%%%%%%%%%%%%%%%%%%%%%%%%%%%%%%%%%%%%%
%%%%%%%%%%%%%%%%%%%%%%%%%%%%%%%%%%%%%%%%%%%%%%%

\section{SUSY breaking in Gepner Models}
\label{sec_gep}

Since the seminal work of D.~Gepner \cite{Gepner:1987qi,Gepner:1987vz}
it is known that there exists a class of
$N=2$ supersymmetric CFTs that describe special points in the moduli
space of Calabi-Yau compactifications. These so-called Gepner models
are the starting point of our construction. It is also known
that the simple current construction can lead to modular invariant
partition functions that break e.g. the left-moving space-time supersymmetry.
In this way one obtains $(0,2)$ or $(1,2)$ superconformal field theories
describing classical ${\cal N}=1$ Minkowski-type string vacua. 
Thus, such special kinds of simple currents provide a CFT realization
of partial supersymmetry breaking.
In this section, we briefly recall the structure of Gepner models
and the simple current construction.  For more details we refer
the reader to the original literature, and for instance to 
\cite{Blumenhagen:2009zz} for a recent review.

\subsection{Review of Gepner construction}

In light-cone gauge, the internal sector of a Type II compactification
to four dimensions with $\mathcal N=2$ supersymmetry is given
by tensor products of the rational models of the $N=2$ super Virasoro
algebra with total central charge $c=9$. 
 Space-time supersymmetry is achieved by  a GSO projection, which can be described
by a certain simple current in the superconformal field theory.

 The minimal models are parametrized by the 
level $k=1,2,\ldots$ and have central charge
\eq{
c=\frac{3\op k }{ k+2}\, .
}
Since $c<3$, the required value $c=9$ is achieved by using  tensor products of 
such minimal models
$ \bigotimes_{j=1}^r (k_j)$. 
The finite number
of irreducible representations of the $N=2$ Virasoro algebra of each
unitary model 
are labeled by the three integers $(l,m,s)$ in the range
\eq{  
\label{range}
l=0,\ldots k\,, \qquad m=-k-1,-k,\ldots k+2\,, \qquad
              s=-1,0,1,2 \,,
}
with $l+m+s=0$ mod $2$. 
Actually, the identification between $(l,m,s)$ and
$(k-l,m+k+2,s+2)$ reveals that the range  \eqref{range} is a double covering of the
allowed representations. 
The conformal dimension and charge of the highest weight 
state with label $(l,m,s)$ are
\eq{
\label{dimensions}
\Delta^l_{m,s}&={l(l+2)-m^2\over 4(k+2)} + \frac{s^2 }{ 8} \,,\\
q^l_{m,s}&={m\over (k+2)}-{s\over 2}  \,.
}
Note that these formulas are  only correct modulo one and two, respectively. 
To obtain the precise conformal dimension $h$ and charge  from \eqref{dimensions}
one  first shifts  the labels into the standard range $|m-s|\le l$
by using the shift symmetries $m\to m+2k+4$, $s\to s+4$ and 
the reflection symmetry.  
The NS-sector consists of those representations with even $s$, while
the ones with odd $s$ are from the R-sector.

In addition to the internal $N=2$ sector, one has the contributions with $c=3$ from the 
two uncompactified
directions. The two world-sheet fermions $\psi^{2,3}$ 
generate a $U(1)_2=SO(2)_1$ model whose four irreducible
representations $(c,o,s,v)$ are labeled by $s_0=-1,\ldots, 2$ with highest weight
and charge modulo one and two respectively
\eq{ \Delta_{s_0}={s_0^2\over 8}  \, ,
             \quad\quad q_{s_0}=-{s_0\over 2} \,.
}
The GSO projection, guaranteeing absence of tachyons and space-time supersymmetry, means in the Gepner case that one projects onto states 
with odd overall $U(1)$ charge $Q_{\rm tot}=q_{s_0}+\sum_{j=1}^r q^{l_j}_{m_j,s_j}$.
Moreover, for having world-sheet supersymmetry with 
\eq{ \label{supercurrent}
G_{\rm tot}=\sum_{j=1}^r  G_j \,+ :\!\partial_z X^\mu \,\psi_\mu\!:\,
} 
beging the overall  $N=1$ world-sheet supercurrent in the product theory,
one has to ensure that in the tensor product only states from the NS respectively 
the R sectors couple among themselves.

These projections are described most conveniently in the following notation.
First one defines some multi-labels
\eq{  \lambda=(l_1,\ldots,l_r)\,, \qquad \mu=(s_0;m_1,\ldots m_r; 
                s_1,\ldots,s_r) \,,
}
and the respective characters 
\eq{ 
\chi^\lambda_\mu(q)=\chi_{s_0}(q)\, \chi^{l_1}_{m_1,s_1}(q)
                        \ldots \chi^{l_r}_{m_r,s_r}(q) \,.
}
In terms of the vectors 
\eq{ 
\beta_0=(1;1,\ldots,1;1,\ldots,1)\,, \hspace{16pt}
            \beta_j=(2;0,\ldots,0;0,\ldots,0,\underbrace{2}_{j^{\rm
                th}},0,\ldots,0)  \,,
}
and the following product
\eq{  Q_{\rm tot}&=2\op \beta_0\bullet \mu =-{s_0\over 2}-\sum_{j=1}^r{s_j\over 2}
              +\sum_{j=1}^r {m_j\over k_j+2}\,,\\
                  \beta_j\bullet \mu &=-{s_0\over 2}-{s_j\over 2} \,,   
}
the projections one has to implement are simply 
$Q_{\rm tot}=2\op\beta_0\bullet \mu\in 2\mathbb Z+1$ and $\beta_j\bullet \mu
\in \mathbb Z$
for all $j=1,\ldots r$.
Gepner has shown that the following GSO projected
partition function
\eq{    Z_D(\tau,\ov{\tau} )={1\over 2^r} 
{ ({\rm Im}\tau)^{-2} \over |\eta(q)|^4 }
     \sum_{b_0=0}^{K-1} \sum_{b_1,\ldots,b_r=0}^1 {\sum_{\lambda,\mu}}^\beta
    (-1)^{s_0} \ \chi^\lambda_\mu (q)\, \chi^\lambda_{\mu+b_0\beta_0
              +b_1 \beta_1 +\ldots b_r \, \beta_r} (\ov q) 
}
is indeed modular invariant and vanishes due to space-time supersymmetry. 
Here $K={\rm lcm}(4,2k_j+4)$ and ${\sum}^\beta$ means that the sum is 
restricted to 
those $\lambda$ and $\mu$ in the range \eqref{range} satisfying 
$2\op\beta_0\bullet \mu\in 2\mathbb Z+1$ and $\beta_j\bullet \mu \in
\mathbb Z$.

\subsection{Simple current extension}

Recall that for a given  conformal field theory there 
exists a very general way to construct  modular invariant partition functions
via  an extension of  the chiral symmetry algebra
by some element of the set of simple currents \cite{Schellekens:1989am,Schellekens:1989dq}.  
These simple currents are primary fields $J_a$ whose operator product expansion
with any other primary field $\phi_i$ only involves one particular other primary 
field, i.e.  
\eq{ J_a \times \phi_i = \phi_{J(i)} 
} 
under fusion. Due to the  associativity of the fusion rules it follows 
that the OPE of two simple currents yields again a simple current,
so that in a rational CFT the set of
simple currents forms a finite abelian group ${\cal S}$ under the fusion product. Being finite there must exist a length ${\cal N}_a$ where ${\cal J}_a^{{\cal N}_a} = 1 $. The set $ \{J_a, J_a^2, \ldots, J_a^{{\cal N}_a} \}$ forms 
an abelian subgroup of ${\cal S}$ isomorphic to  ${\mathbb Z}_{{\cal N}_a}$ with  
$ (J_a^n)^C \equiv (J_a ^{ n}) ^{-1} = J_a ^{{\cal N}_a - n}$. 
Similarly, every simple current groups the primary fields into orbits 
$\{ \phi_i, J_a \times \phi_i, J_a^2 \times \phi_i, 
\ldots, J_a^{{\cal N}^i_a - 1}\times \phi_i\}$ whose length ${\cal
N}^i_a$ is a divisor of ${\cal N}_a$.
 
The crucial observation  is that the action of simple currents in a 
RCFT implies the existence of a conserved quantity for every
primary $\phi_i$, the monodromy charge $Q^{(a)}_i$, defined by
\eq{ 
J_a(z) \,  \phi_i(w) = (z - w )^{ - Q^{(a)}_i} \phi_{J(i)} (w)+\ldots \,. 
}
The monodromy of the identity being 1, it is clear that 
$ Q^{(a)}_i = {t^i_a\over {\cal N}_a}\ {\rm mod}\ 1$ for some integer $t^i_a$.
On the other hand, the monodromy is given by the conformal dimensions of the primary and 
the simple current as
\eq{ 
Q^{(a)}_i = h (\phi_i) + h ( J_a) - h (J_a \times \phi_i) \ {\rm mod}\
1\,, 
}
from which one can derive that 
\eq{ 
Q ^{(a)}_i(J_a^n \times \phi_i) = {t^i_a + r_a n  \over {\cal N}_a} \
{\rm mod}\ 1\,.
} 
Here the monodromy parameter $r_a$ is defined such that 
\eq{
h (J_a) = {r_a ({\cal N}_a - 1 ) \over 2{\cal N}_a}  \ {\rm mod}\  1\,.
}
One can prove that 
a simple current $J_a$ with even monodromy parameter $r_a$ induces the 
 modular invariant partition function
\eq{
Z_a ( \tau, \bar{\tau})  =  \vec\chi^{\,T}(\tau)\, M(J_a)\,\vec\chi(\ov\tau)=
\sum_{k,l} \chi_k (\tau)\,\, (M_a)_{kl}\,\,
\chi_l (\bar{\tau})\,,
} 
where 
\eq{
(M_a)_{kl} = \sum_{p = 1}^{{\cal N}_a} \delta( \phi_k, J_a^p \times \phi_l) \, \, \, 
\delta^{(1)} \big(\hat{Q}^{(a)}(\phi_k) + \hat{Q}^{(a)} (\phi_l) \big) 
}
and 
\eq{
\hat{Q}^{(a)} (\phi_i) = {t_a^i \over 2 {\cal N}_a} \ {\rm mod}\ 1 \,. 
}
Note that the proof relies on the fact that $r_a$ is even, which can always be arranged 
for odd ${\cal N}_a$, with $r_a$ being defined only ${\rm mod}\, {\cal N}_a$.

Given two modular invariant matrices 
$M_{a_1}$ and $M_{a_2}$, it is clear that
$Z_{a_1, a_2} = {1\over N}\sum_{k,l,m} \chi_l \, (M_{a_1})_{lk} \, (M_{a_2})_{km}\, \chi_m$ 
is another modular invariant partition function with obvious generalizations for 
several $M_{a_i}$; 
the normalization factor $N$ ensures
that the vacuum appears precisely once in $Z_{a_1, a_2}$.  The matrices $M$ are
also seen to  commute if the respective simple 
currents $J_{a_1}$ and $J_{a_2}$ are mutually local, i.e. if their 
relative monodromy charge $Q^{(a_1)} (J_{a_2}) = 0  \ {\rm mod}\  1$.

\subsection{Asymmetric Gepner models}

In order to directly apply the simple current extension to the 
Gepner model, one needs to apply the bosonic string map
that exchanges $SO(2)_1\to SO(10)_1\otimes (E_8)_1$ and maps the
four representations as
\eq{
          \phi_{\rm bsm}: (o,v,s,c)\to (v,o,-c,-s)\otimes 1 \,.
}  
In this way one obtains a purely bosonic CFT without any minus signs
in the modular invariant partition function.
Given the fusion rules
\eq{
\phi_{ (m_1, s_1)}^0 \times \phi_{ (m_2, s_2)}^{l_2}  =  
          \phi_{ (m_1 + m_2, s_1 + s_2)} ^{l_2} \,,
}
we conclude that the simple currents $J$ of the Gepner model under 
consideration can be labeled by the vector
\eq{
J=(0\; m_1\; s_1)\ldots (0\; m_5\; s_5)(s_0)\,.
}
The Gepner partition function can then be expressed as the simple current
extension
\eq{ \label{partitionfn}
              Z_{\rm Gepner}(\tau,\ov\tau)\sim \vec\chi^{\,T}(\tau)\, M(J_{\rm GSO})\,
              \prod_{r=1}^5 M(J_i) \,\vec\chi(\ov\tau)\Big\vert_{\phi^{-1}_{\rm bsm}}
}
with the bosonic string map applied backwards at the end.
The simple currents are given by
\eq{ \label{simplecurrents}
              J_{\rm GSO}&=(0\;1\; 1)\ldots (0\; 1\; 1)(s)\,, \\[0.1cm]
             J_i&=(0\;0\; 0)\ldots \underbrace{(0\;0\; 2)}_{i^{\rm
                th}}\ldots (0\; 0\; 0)(v) \,.
}
The $J_i$ are also  called alignment simple currents.
Having a minimal model with even level e.g. in the last factor one can also use its D-type modular invariant by adding the simple current
\eq{ \label{Dsimplecurrent}
J_D = (0 \; 0 \; 0)\ldots  (0 \; 0 \;0) (0 \; k+2 \;2)  (o)
}
in the partition function. All these simple currents above are relatively local to each other.

The massless spectrum can be read off from the partition function
and consists of the ${\cal N}=2$ supergravity multiplet, the universal 
hypermultiplet, $N_V$ vectormultiplets and additional $N_H-1$
hypermultiplets. For the Type IIB superstring this CFT describes a special point
in the moduli space of a Calabi-Yau compactification
with Hodge numbers $h^{2,1}=N_V$ and $h^{1,1}=N_H-1$. 
Note that the supergravity multiplet contains the graviphoton and
the universal hypermultiplet contains the dilaton and the four-dimensional 
Kalb-Ramond field $B_{\mu\nu}$.

As exploited in the heterotic context in \cite{Schellekens:1989wx,Blumenhagen:1995tt} and for Type II in
\cite{Israel:2013wwa,Israel:2015efa}, the method of simple currents is very well suited to
construct asymmetric CFTs. One only has to further extend
the Gepner partition function by a simple current $J_{\rm ACFT}$ that 
is not local with respect to $J_{\rm GSO}$ and/or $J_i$. Indeed in the
partition function
\eq{
\label{gepnerasym}
              Z_{\rm ACFT} (\tau,\ov \tau)\sim \vec\chi^{\,T}(\tau)\, M(J_{\rm ACFT})\,
              M(J_{\rm GSO})\,
              \prod_{r=1}^5 M(J_i) \,\vec\chi(\ov\tau)\Big\vert_{\phi^{-1}_{\rm bsm}} 
}
the spectral flow (left-moving supercharge) does not act on the
left-moving side leading to an asymmetric CFT-model with only ${\cal N}=1$ space-time
supersymmetry. 

If $J_{\rm ACFT}$ does not commute with $J_{\rm GSO}$ then the
left-moving  space-time supersymmetry is broken. For interpreting the
model as a type II compactification, it was so 
far assumed \cite{Schellekens:1989wx,Israel:2013wwa,Israel:2015efa}
that $J_{\rm ACFT}$ has to commute with the simple currents $J_i$, that
were implemented in order to realize $N=1$ supersymmetry on the tensor
product with the supercurrent given by \eqref{supercurrent}. This means that
$J_{\rm ACFT}$ should contain only  NS or R entries for each tensor
factor.
For our purposes we will not require this strong condition, as
from the point of view of the central charge $c=(12,12)$, modular
invariance, absence of tachyons and
the arising multiplet structure of the massless spectrum we do not see any difference.
Indeed we think that this mixing of NS and R entries in $J_{\rm ACFT}$
does not necessarily break the left-moving $N=1$ world-sheet
supersymmetry. As we will see in section \ref{sec_oddlev},   in this class we can construct
models with even ${\cal N}=2$ space-time supersymmetry.

In any case, by construction \eqref{gepnerasym}  is a classical
tachyon-free string vacuum with the central charge $(12,12)$ that 
has  vanishing cosmological constant, i.e. it is a Minkowski vacuum.
Of course one can add more simple currents, but in this paper we only
consider the simplest case with only a single one. The massless spectrum still fits into ${\cal N}=1$ supermultiplets.
From the vacuum orbit one gets the ${\cal N}=1$ supergravity multiplet
and a chiral superfield containing the dilaton and another pseudo-scalar.
The graviphoton is not massless anymore. From the so-called matter orbits, one gets three kinds of
massless states.
Left-right combinations of the form
\eq{
        (h=3/8)(s)  \otimes  \Big[ (\ov h=1/2,\ov q=1)(o)+(\ov  h=3/8,\ov q=-1/2)(s)\Big]
}
lead to massless ${\cal N}=1$ vectormultiplets, combinations
\eq{
        (h=3/8)(c)  \otimes   \Big[(\ov h=1/2,\ov q=1)(o)+(\ov
        h=3/8,\ov q=-1/2)(s) \Big]
}
to massless R-R axion-like chiral multiplets and
\eq{
        (h=1/2)(o)  \otimes  \Big[ (\ov h=1/2,\ov q=1)(o)+(\ov  h=3/8,\ov q=-1/2)(s)\Big]
}
to NS-NS scalar chiral multiplets.
Therefore, the latter three classes of  massless states are described by three numbers
$(N_V,N_{\rm ax};N_0)$.

In section~\ref{sec_examples} we will discuss a couple of examples for
$J_{\rm ACFT}$. All these are very simple, in the sense that they
trivially act on most of the tensor factors.
Recall that it is the aim of this paper to provide
arguments for the identification of  asymmetric Type IIB
Gepner models \eqref{gepnerasym} with ${\cal N}=1$ Minkowski vacua of
${\cal N}=2$ gauged supergravity models.
Therefore, let us first recollect  the structure of ${\cal N}=2$ GSUGRA and its
partial supersymmetry breaking vacua.

%%%%%%%%%%%%%%%%%%%%%%%%%%%%%%%%%%%%%%%%%%%%%%%
%%%%%%%%%%%%%%%%%%%%%%%%%%%%%%%%%%%%%%%%%%%%%%%
%%%%%%%%%%%%%%%%%%%%%%%%%%%%%%%%%%%%%%%%%%%%%%%
%%%%%%%%%%%%%%%%%%%%%%%%%%%%%%%%%%%%%%%%%%%%%%%
%%%%%%%%%%%%%%%%%%%%%%%%%%%%%%%%%%%%%%%%%%%%%%%
%%%%%%%%%%%%%%%%%%%%%%%%%%%%%%%%%%%%%%%%%%%%%%%

\section{Partial SUSY-breaking in $\mathcal N=2$ GSUGRA}
\label{sec_sugra}

In this section we review some aspects of $\mathcal N=2$ gauged supergravity  
theories and their partial breaking to $\mathcal N=1$, for which we mostly follow
\cite{Louis:2009xd,Louis:2010ui,Hansen:2013dda}.
We consider GSUGRAs
resulting from flux compactifications of type IIB superstring theories, which 
are related to abelian gaugings along the axionic directions
in the hypermultiplet moduli space.

%%%%%%%%%%%%%%%%%%%%%%%%%%%%%%%%%%%%%%%%%%%%%%%
%%%%%%%%%%%%%%%%%%%%%%%%%%%%%%%%%%%%%%%%%%%%%%%
%%%%%%%%%%%%%%%%%%%%%%%%%%%%%%%%%%%%%%%%%%%%%%%
%%%%%%%%%%%%%%%%%%%%%%%%%%%%%%%%%%%%%%%%%%%%%%%

\subsection{Basics of $\mathcal N=2$ GSUGRA}

We begin by reviewing $\mathcal N=2$ gauged supergravity theories in four dimensions,
which arise from Calabi-Yau compactifications of superstring theory with fluxes. 
The field content in the four-dimensional theory is given by one supergravity multiplet, $N_V$ vector-multiplets and $N_H$ hyper-multiplets. 
For type IIB string theory we have $N_V=h^{2,1}$ and $N_H=h^{1,1}+1$.
The multiplets contain the following degrees of freedom
\eq{
  \label{multiplets_01}
  \arraycolsep2pt
  \renewcommand{\arraystretch}{1.5}
  \begin{array}{@{}l@{\hspace{25pt}}lclcl@{}}
  \mbox{massless $\mathcal N=2$ gravity}&\mathcal G_{(2)} &=& 1\cdot [2] + 2\cdot [\tfrac{3}{2}] + 1\cdot [1] 
  &=& (2)_{\rm b} + (4)_{\rm f} + (2)_{\rm b} \,, \\
  \mbox{massless $\mathcal N=2$ vector} &\mathcal V_{(2)} &=& 1\cdot [1] + 2\cdot [\tfrac{1}{2}] + 2\cdot [0] 
  &=& (2)_{\rm b} + (4)_{\rm f} + (2)_{\rm b} \,, \\
  \mbox{massless $\mathcal N=2$ hyper} &\mathcal H_{(2)} &=& 2\cdot [\tfrac{1}{2}] + 4\cdot [0] 
  &=& (4)_{\rm f} + (4)_{\rm b} \,,
  \end{array}
}
where the number in a square bracket indicates the spin and the number in parenthesis 
counts the real bosonic and fermionic degrees of freedom. 
Compactification of type IIB string theory on Calabi-Yau manifolds are well-understood,
and in the following we therefore only summarize the features needed here.

%%%%%%%%%%%%%%%%%%%%%%%%%%%%%%%%%%%%%%%%%%%%%%%
%%%%%%%%%%%%%%%%%%%%%%%%%%%%%%%%%%%%%%%%%%%%%%%

\subsubsection*{NS-NS  sector}

One introduces a symplectic 
basis for the third cohomology group of the Calabi-Yau three-fold $\mathcal M$ as 
$\{\alpha_{\Lambda},\beta^{\Lambda}\} \in H^3(\mathcal M)$ 
with $\Lambda =0,\ldots, h^{2,1}$.
The holomorphic three-form $\Omega$ 
can be expanded as
\eq{
\label{hol_three}
\Omega=X^\Lambda\, \alpha_\Lambda - F_\Lambda\, \beta^\Lambda =
\bigl( \alpha_{\Lambda}\,,\,\beta^{\Lambda}\bigr) \cdot V_2 \,,
} 
where we follow the conventions in \cite{D'Auria:2007ay,Blumenhagen:2015lta} and 
introduced a $(2\op h^{2,1}+2)$-dimensional vector as
 $V^T_2 = (X^{\Lambda}\,,\, -F_{\Lambda} )$.
The periods $X^{\Lambda}$ are projective coordinates on the moduli space,
and are related to the complex-structure moduli through
$z^a = X^a/X^0$ where $a=1,\ldots, h^{2,1}$.
We also mention that it is usually assumed that the periods $F_{\Lambda}$ 
can be written as the derivative of a prepotential $F(X)$ with respect to $X^{\Lambda}$, that 
is $F_{\Lambda} = \partial  F/\partial X^{\Lambda}$.
For later reference we furthermore define the invertible and positive-definite matrix $\mathcal M_1$, 
which can be expressed in terms of the period matrix $\mathcal N$ as
\eq{
\label{res_016}
    {\cal M}_1=\left(\begin{matrix} \mathds 1 & {\rm Re}\,{\cal N} \\ 0 &
        \mathds 1 \end{matrix}\right) 
     \left(\begin{matrix} - {\rm Im}\,{\cal N} & 0 \\ 0 &
        - {\rm Im}\,{\cal N}^{-1}  \end{matrix}\right)
   \left(\begin{matrix} \mathds 1 & 0\\ {\rm Re}\,{\cal
         N}   &
        \mathds 1 \end{matrix}\right) \,.
}

For the even cohomology of the Calabi-Yau manifold $\mathcal M$  
one finds a similar special geometry. 
We introduce bases of the form
$\{ \omega_{\lab A} \}  \in  H^{1,1}(\mathcal M)$
and
$\{ \sigma^{\lab A} \}  \in  H^{2,2}(\mathcal M)$
with $\lab A = 1,\ldots, h^{1,1}$.
We can group these two- and four-forms together with the zero- and six-form as
$ \{ \omega_{ A} \}  =  \{ 1,\, \omega_{\lab A} \bigr\}$
and   $\{ \sigma^{ A} \} = \bigl\{ \tfrac{\sqrt{g}}{{\cal V}} \op dx^6,\,\sigma^{\lab A}\bigr\} $ with 
$A = 0,\ldots, h^{1,1}$.
Here $\mathcal V$ is the volume of $\mathcal M$. 
The K\"ahler form $J$ of  $\mathcal M$
and the Kalb-Ramond field $B$ are expanded in the basis $\{\omega_{\lab A}\}$ in the following way
\eq{
J=t^{\lab A} \op \omega_{\lab A} \,, \hspace{70pt}
B = b^{\lab A}\op\omega_{\lab A}\,,
}
and can be combined into a complex field as $\mathcal J= B + i J= \mathcal J^{\lab A} \op\omega_{\lab A}$.
Note that $\mathcal J^A$ are the $h^{1,1}$ complexified K\"ahler moduli.
We 
introduce a complex $(2h^{1,1}+2)$-dimensional vector $V_1$ as
\eq{
     e^{B+i J}=e^{\mathcal J} = \bigl( \omega_A\,,\,\sigma^A\bigr) \cdot V_1 \,,
}
where the components of $V_1$ read
\eq{
  V_1 = \left( \begin{array}{c}
  1 \\
  \mathcal J^{\lab A} \\
  \tfrac{1}{6}\op \kappa_{\lab{ABC}} \mathcal J^{\lab A} \mathcal J^{\lab B} \mathcal J^{\lab C}\\
   \tfrac{1}{2}\op \kappa_{\lab{ABC}} \mathcal J^{\lab B} \mathcal J^{\lab C} 
  \end{array}
  \right) .
}
Finally, in analogy to \eqref{res_016} there exists a positive definite
and invertible matrix ${\cal M}_2$.
The precise expressions are not important here, but can be found 
for instance in section 4.1 of \cite{Blumenhagen:2015lta}. 

%%%%%%%%%%%%%%%%%%%%%%%%%%%%%%%%%%%%%%%%%%%%%%%
%%%%%%%%%%%%%%%%%%%%%%%%%%%%%%%%%%%%%%%%%%%%%%%

\subsubsection*{R-R sector}

The Ramond-Ramond sector of type IIB provides additional massless modes, 
that will play  the dominant role in our investigation. 
The four-dimensional scalar part of the R-R potentials is obtained as follows
\eq{
  \mathcal C \,\Bigr\rvert_{\rm scal.}= 
  \tilde \xi_0+
  \xi^{\lab A} \op \omega_{\lab A} +
  \tilde\xi{}_{\lab A} \op \sigma^{\lab A} +  
  \xi^0\op \omega_0 
  = \bigl( \omega_A\, ,\,\sigma^A\bigr) \cdot \Xi
  \,,
  \hspace{30pt}
  \Xi = \binom{\xi^A}{\tilde\xi_A} \,,
}
where $\mathcal C=C_0+C_2+C_4+C_6+C_8$ is a formal sum of R-R forms
in type IIB.
This expansion defines a $(2\op h^{1,1} + 2)$-dimensional vector $\Xi$
of R-R axions.
The pairs $( \xi^{\lab A},\tilde\xi_{\lab A})$  
form $h^{1,1}$ complex axionic scalars, which pair up with  the
complexified K\"ahler moduli ${\cal J}^{\lab A}$ to form 
$h^{1,1}$ hyper-multiplets. The two remaining
R-R axions $( \xi^0,\tilde\xi_0)$ combine with the dilaton $\phi$ 
and the NS-NS axion $\tilde \phi$ to the so-called universal hypermultiplet.
The axion $\tilde\phi$ is the dual to the four-dimensional Kalb-Ramond field $B_{\mu\nu}$.

Turning now to the four-dimensional vector fields, we expand
\eq{
 C_4= A^{\Lambda} \op\alpha_{\Lambda} +   \tilde A_{\Lambda} \op \beta^{\Lambda} \,,
}
in which $(A,\tilde A)$ are four-dimensional electric and magnetic vector fields. Eventually,
half of these have to be eliminated due to the self-duality condition on 
the R-R five form, leaving only $h^{2,1}+1$ vectors. 
But for now we will keep them as separate degrees of freedom. 
Here the vector field $A^0$ is the graviphoton residing in the ${\cal
  N}=2$
supergravity multiplet and the remaining $h^{2,1}$ gauge fields combine
with the complex structure moduli to fill out $h^{2,1}$
vector-multiplets.
Thus, the bosonic components of the ${\cal N}=2$ supergravity  multiplets are  
\eq{
  \arraycolsep2pt
  \renewcommand{\arraystretch}{1.5}
  \begin{array}{l@{\hspace{55pt}}lcl}
  \mbox{massless $\mathcal N=2$ gravity}&\mathcal G_{(2)} &\supset& \bigl(\op g_{\mu\nu}\, ,\: 
  A^0 \bigr) \,, \\
  \mbox{massless $\mathcal N=2$ vector} &\mathcal V_{(2)} &\supset& \bigl( \, A^a\,, \: z^a \bigr) \,, \\
  \multirow{2}{*}{\mbox{massless $\mathcal N=2$ hyper}}
   &\mathcal H_{(2)} &\supset& \bigl(  \mathcal J^{\lab A}\,,\:
  \xi^{\lab A}\,,\: \tilde\xi_{\lab A}\bigr) \,, \\
  &\mathcal H^{\rm univ.}_{(2)} &\supset& \bigl(  \phi \,,\: \tilde\phi \,,\:
  \xi^{0}\,,\: \tilde\xi_0\bigr) \,. \\
  \end{array}
}
The complex structure moduli $z^a$ are coordinates  on a special K\"ahler manifold.
The $4\op(h^{1,1}+1)$ scalars in the hypermultiplets form a
special hyper-K\"ahler manifold which is a fibration of dimension $2h^{1,1}+4$
over a special K\"ahler manifold described by the $h^{1,1}$ complex K\"ahler
moduli $\mathcal J^{\lab A}$.

%%%%%%%%%%%%%%%%%%%%%%%%%%%%%%%%%%%%%%%%%%%%%%%
%%%%%%%%%%%%%%%%%%%%%%%%%%%%%%%%%%%%%%%%%%%%%%%

\subsection{Gaugings via background fluxes}

We now turn to compactifications on Calabi-Yau three-folds with background
fluxes, which lead to gaugings of (abelian) hyper-multiplet isometries
\cite{Grana:2005ny,Louis:2002ny,Dall'Agata:2004nw,D'Auria:2004tr,D'Auria:2004wd,D'Auria:2007ay}.
The fluxes  can be geometric in the NS-NS sector ($H$-flux and geometric flux)
or R-R sector ($F^{(3)}$-flux), or non-geometric in the NS-NS sector
($Q$- and $R$-flux).

Concerning  the R-R sector of type IIB string theory, the three-form flux $F^{(3)}$, 
can be expanded  as 
\eq{
  \label{res_030}
  F^{(3)} = -\tilde{\mathsf F}^{\Lambda} \op\alpha_{\Lambda} + {\mathsf F}_{\Lambda} \op \beta^{\Lambda}
  =\bigl( \alpha_{\Lambda}\hspace{4pt}\beta^{\Lambda}\bigr) \cdot \mathsf F^{(3)} \,,
  \hspace{50pt}
  \arraycolsep1.5pt
  \mathsf F^{(3)} = \left(\begin{array}{r}-\tilde{\mathsf F}^{\Lambda} \\ {\mathsf F}_{\Lambda} \end{array}
  \right)\,.
}
The geometric and non-geometric NS-NS fluxes are conveniently organized into a
$(2\op h^{2,1}+2)\times (2\op h^{1,1}+2)$ matrix
as follows (see e.g. \cite{Grana:2006hr})\footnote{As compared to \cite{Blumenhagen:2015lta}, 
we changed our conventions for the fluxes as $\tilde{\mathcal O}_{\rm there} = -\mathcal O_{\rm here}$.}
\eq{
  \label{res_015}
{\cal O}=\left(\begin{array}{@{\hspace{2pt}}ll@{\hspace{2pt}}}
  q_{\Lambda}{}^{ A} & f_{\Lambda \op A} \\[4pt]
  \tilde q^{\Lambda\op  A} & \tilde f^{\Lambda}{}_ { A}
\end{array}\right).
}
Note that the $H$- and $R$-flux are contained in \eqref{res_015} via
\eq{
\label{fluxzerocomp}
\arraycolsep2pt
\begin{array}{lcl@{\hspace{70pt}}lcl}
f_{\Lambda \op0}&=&h_\Lambda\,, & \tilde f^{\Lambda}{}_0&=&\tilde h^\Lambda\,,\\[5pt]
q_{\Lambda}{}^0&=&r_\Lambda\,,& \tilde q^{\Lambda\op 0}&=&\tilde r^\Lambda\, .
\end{array}
}
These fluxes lead to a gauging of isometries in the hypermultiplet
moduli space. More concretely, the shifts along the $2h^{1,1}+3$ axionic
directions $\{\xi^A,\tilde\xi_A, \tilde\phi\}$ are gauged according to
\eq{
  \label{gauged}
  \delta  \left(\begin{array}{@{\hspace{2pt}}r@{\hspace{2pt}}}A\\
      {\tilde  A} \end{array}
  \right) = d\op\lambda\,,
  \hspace{20pt}
  \delta\op \Xi = -\mathcal O^T\cdot \lambda \,,
  \hspace{20pt}
  \delta\op \tilde\phi = -2\op \lambda^T \cdot \mathsf F^{(3)} - \lambda^T \cdot C\cdot \tilde{\mathcal O} \cdot \Xi \,,
}
where $\lambda$ is a $(2\op h^{2,1}+2)$-dimensional vector  parametrizing the gauge transformation,
and where we have defined the matrices
\eq{
\tilde{\mathcal O} = C\cdot \mathcal O\cdot C^T \,,
\hspace{30pt}    C=\left(\begin{matrix} 0 & +\mathds 1 \\ -\mathds 1 &
    0\end{matrix}\right),
}
with the dimensions of the square matrix $C$ chosen appropriately.
In this notation the quadratic constraints (Bianchi identities) for
the fluxes can be expressed as 
\eq{
  \label{cons_01}
  \tilde{\mathcal O}^T \cdot {\mathcal O} = 0 \,,
  \hspace{40pt}
  {\mathcal O}\cdot \tilde{\mathcal O}^T = 0 \,.
}
Through such a gauging $n={\rm rank}({\mathcal O})+\Delta$ gauge fields become
massive via the St\"uckelberg mechanism 
by eating some of the axions. The extra contribution $\Delta\in\{0,1\}$ 
is equal to one if the NS-NS axion $\tilde\phi$ is gauged as well.
(For  details on the gauging of $\tilde\phi$ see \cite{Louis:2002ny}.)
Moreover, the gaugings induce  a scalar potential that in general
depends on all types of moduli and is given by \cite{D'Auria:2007ay,Cassani:2008rb}
\eq{
  \label{res_022}
  V = &\hspace{14pt}
  \frac{1}{2} \op \bigl( \mathsf F^T - \Xi^T \cdot \tilde{\mathcal O}^T \bigr) \cdot \mathcal M_1 \cdot
  \bigl( \mathsf F -\tilde{ \mathcal O} \cdot \Xi \bigr) \\
  & + \frac{e^{-2\phi}}{2} \op V_1^T\cdot \tilde{\cal O}^T \cdot {\cal M}_1\cdot \tilde{\cal O}\cdot \ov V_1\\
  & + \frac{e^{-2\phi}}{2} \op   V_2^T \cdot {\mathcal O} \cdot \mathcal M_2 \cdot {\mathcal O}^T \cdot \ov V_2
  \\
  &-\frac{e^{-2\phi}}{4\op \mathcal V} \, 
  V_2^T \cdot C \cdot \tilde{\mathcal O}\cdot \Bigl(  V_1 \times \ov V_1^T + \ov V_1 \times  V_1^T  \Bigr)
   \cdot \tilde{\mathcal O}^T\cdot C^T \cdot \ov V_2 \,.
}
Note that the R-R axions $\Xi$ only appear in the first term and that  
the scalars gauged via  \eqref{gauged} do not appear in the scalar potential \eqref{res_022}. Indeed, 
due to 
\eq{
  \delta_{\lambda} \bigl( \tilde{\mathcal O}\cdot \Xi\bigr) = -\tilde{\mathcal O} \cdot {\mathcal O}^T\cdot \lambda= 0\,,
}
the scalar potential is gauge invariant. Furthermore, $\tilde\phi$ does not appear in \eqref{res_022}.

To summarize, $n$  of the R-R
axions $\Xi$ and the NS-NS axion $\tilde\phi$ can become massive via  the St\"uckelberg
mechanism, while the remaining axions can still receive a mass from the
scalar potential. However, the axions only appear via the combination
$\tilde{\cal O}\cdot \Xi$ which for $h^{2,1}>h^{1,1}$  can be shown to depend only
on $h^{11}+1$ combinations of axions. Therefore, at most $h^{1,1}+1$
can receive a mass from the potential. We will see in the next section
that for supersymmetric minima, this upper bound is actually smaller.

Conceptually, the scalar potential \eqref{res_022} can be obtained from a
dimensional reduction of double field theory (DFT) on a Calabi-Yau manifold
equipped with NS-NS and R-R fluxes, where the latter are treated as small
perturbations around the CY geometry \cite{Blumenhagen:2015lta}. 
Furthermore, from a supergravity point of view, \eqref{res_022} corresponds to 
$SU(3)\times SU(3)$ structure compactifications 
\cite{Grana:2005sn,Grana:2005ny,Benmachiche:2006df,Grana:2006hr,Cassani:2007pq}.
However, it is not clear whether four-dimensional GSUGRA can be considered as a low-energy 
effective action (LEEA) 
for the light modes in a string compactification.
First, even DFT itself is rather a truncation of string theory than an
LEEA and second, having non-geometric fluxes turned on implies that in general there does not exist 
a dilute flux limit for which the backreaction of the fluxes on the CY 
can be argued to be small\cite{Blumenhagen:2015kja}. 
It is thus not clear  whether minima
of the scalar potential of GSUGRA can be truly uplifted to full
classical solutions of the string equations of motion.

%%%%%%%%%%%%%%%%%%%%%%%%%%%%%%%%%%%%%%%%%%%%%%%
%%%%%%%%%%%%%%%%%%%%%%%%%%%%%%%%%%%%%%%%%%%%%%%
%%%%%%%%%%%%%%%%%%%%%%%%%%%%%%%%%%%%%%%%%%%%%%%
%%%%%%%%%%%%%%%%%%%%%%%%%%%%%%%%%%%%%%%%%%%%%%%

\subsection{Partial supersymmetry breaking}

We now briefly  describe spontaneous supersymmetry breaking from $\mathcal N=2$ to 
$\mathcal N=1$ following the work of
\cite{Louis:2009xd,Louis:2010ui} (see also \cite{Andrianopoli:2002vq,Cassani:2009na}).
As has been shown by these authors, such a breaking is possible if magnetic gaugings 
and non-geometric fluxes are considered.

Our goal in this section is to deduce bounds on  the number of massless  vector- and
R-R chiral multiplets in ${\cal N}=1$ Minkowski vacua of ${\cal N}=2$
GSUGRA. 
In the spontaneously-broken theory the following multiplets are of importance, which we 
summarize using the same notation as in \eqref{multiplets_01}:
\eq{
  \arraycolsep2pt
  \renewcommand{\arraystretch}{1.5}
  \begin{array}{@{}l@{\hspace{25pt}}lclcl@{}}
  \mbox{massless $\mathcal N=1$ gravity} &G_{(1)}&=& 1\cdot [2] + 1\cdot [\tfrac{3}{2}] 
  &=& (2)_{\rm b} + (2)_{\rm f}\,,  \\
  \mbox{massless $\mathcal N=1$ vector} &V_{(1)}&=& 1\cdot [1] + 1\cdot [\tfrac{1}{2}] 
  &=& (2)_{\rm b} + (2)_{\rm f} \,, \\
  \mbox{massless $\mathcal N=1$ chiral} &C_{(1)}&=& 1\cdot [\tfrac{1}{2}] + 2\cdot [0] 
  &=& (2)_{\rm f} + (2)_{\rm b} \,, \\[10pt]
  \mbox{massive $\mathcal N=1$ spin-$3/2$} &\ov S_{(1)}&=& 1\cdot [\tfrac{3}{2}] + 2\cdot [1] + 1\cdot [\tfrac{1}{2}]
  &=& (4)_{\rm f} + (6)_{\rm b} + (2)_{\rm f} \,, \\
  \mbox{massive $\mathcal N=1$ vector} &\ov V_{(1)}&=& 1\cdot [1] + 2\cdot [\tfrac{1}{2}] + 1\cdot [0]
  &=& (3)_{\rm b} + (4)_{\rm f} + (1)_{\rm b} \,, \\
 \mbox{massive $\mathcal N=1$ chiral} &\ov C_{(1)}&=& 1\cdot [\tfrac{1}{2}] + 2\cdot [0]
  &=& (2)_{\rm f} + (2)_{\rm b} \,.  \\
  \end{array}
}

The breaking mechanism can be separated into two steps. 
The first step is responsible for the partial supersymmetry breaking, 
in which one gravitino of the $\mathcal N=2$ gravity-multiplet becomes massive
while the other stays massless. 
The latter will be part of the $\mathcal N=1$ gravity multiplet $G_{(1)}$, while the 
former is part of a massive spin-$3/2$ multiplet $\ov S_{(1)}$. 
Since the broken theory is required to be $\mathcal N=1$ supersymmetric, 
the massive spin-$3/2$ multiplet has to contain two massive 
vector fields, which acquire a mass from the St\"uckelberg  mechanism by eating 
two gauged axions. The axions can -- but do  not have to -- include
the NS-NS field $\tilde\phi$.

Furthermore, for an ${\cal N}=1$ vacuum with vanishing R-R flux, 
axions $\zeta_{1,2}$ are generically fixed by the complex valued relation
\eq{ \label{RRfixing}
                    ( \tilde\xi_A-{G}_{AB}\, \xi^B)\, D^A=0\,,
}
where ${G}_{AB}=\partial_A\partial_B {G}$ with ${G}$
denoting the prepotential for  the K\"ahler moduli space. $D^A$ is a
constant complex valued vector that specifies the ${\cal N}=1$ vacuum
(see \cite{Louis:2009xd,Louis:2010ui} for more details). Since the
axions only appear quadratically in the scalar potential \eqref{res_022}, these two axions 
will receive a mass. However, one can imagine that for some
boundary values in the K\"ahler moduli space ${G}_{AB}$ degenerates
such that only one axion is fixed. Thus, in the following 
we only assume that at least one axion is fixed by \eqref{RRfixing}.
Recalling then that we have $h^{2,1}+1$ gauge fields to begin with and
$2(h^{1,1}+1)+1$ real axions where only one is from the NS-NS sector,
after this first step we are generically left with $N_V=h^{2,1}-1$ massless vectors and 
$N^{\rm real}_{\rm ax}=2h^{1,1}+1-k_1$ massless real axions. 
Here $k_1\in\{1,2\}$ denotes the number of axions that became massive
due to \eqref{RRfixing}.

In the second step, there can be additional gaugings which however do
not participate in the partial supersymmetry breaking and can
therefore be described in an effective ${\cal N}=1$ GSUGRA theory.
For $n-2$ additional gaugings, $n-2$ vector fields become massive
eating $n-2$ axions. In addition one gets an F-term and
a D-term potential, where the axions only appear in the D-terms for
the broken abelian gauge fields.
In an ${\cal N}=1$ SUSY preserving Minkowski minimum the D-terms have
to vanish providing  $n-2$ real conditions\footnote{In \cite{Hansen:2013dda},
these D-term conditions  were given by 
\eq{
       {\rm Re}\left( (s_{\lambda A}- r_\lambda{}^C\, G_{CA})\, ({\rm Im}G^{-1})^{AB}\,      
                            (\tilde\xi_B-\overline G_{BD} \xi^D) - t_\lambda
                          \right)=0
}
where $s_{\lambda A}, r_\lambda{}^A$ and $t_\lambda$ denote the
components of the Killing vectors that are gauged. Note that these
conditions depend on the axions and the K\"ahler moduli.
}. Therefore, at most $n-2$
additional axions can  become massive by this mechanism. The remaining axions
are free parameters and therefore massless. In summary,  we finally 
get $N_V=h^{2,1}-n+1$ massless vectors
and $N^{\rm real}_{\rm ax}=2h^{1,1}-n+3-(k_1+k_2)$ real massless axions with
$0\le k_2\le n-2$ denoting the number of  axions that became massive
due to the $n-2$ D-terms.

Taking into account the maximal number of gaugings, we have the bound $2\le n\le
h^{1,1}+1+\Delta$, 
where the extra term $\Delta$ appears only for a  gauging along the NS-NS axion $\tilde\phi$.
Therefore, for the number of vector multiplets after gauging we can
derive the bound
\eq{
\label{relav}
                      h^{2,1}-h^{1,1}-\Delta\le N_V\le h^{2,1}-1\,.
}
Moreover, using $n=h^{2,1}-N_V+1$, for the number of real massless
axions we find
\eq{
            2(h^{1,1}-h^{2,1}+N_V)+1\le   N^{\rm real}_{\rm ax}\le 2h^{1,1}-h^{2,1}+N_V+1 \,.
}
If the $\tilde\phi$-field is gauged then the number of complex R-R axions is
$N_{\rm ax}=N^{\rm real}_{\rm ax}/2$, whereas for an ungauged $\tilde\phi$ one
has $N_{\rm ax}=(N^{\rm real}_{\rm ax}-1)/2$. Thus, for the number of
complex R-R axions we derive the bounds
\eq{
\label{relaax1}
              N_V-N_{\rm ax}\le h^{2,1}-h^{1,1}-\Delta \,,
}
and
\eq{
\label{relaax2}
      N_V-2N_{\rm ax}\ge h^{2,1}-2h^{1,1}-\Delta\,.
}
These numbers are to  be compared to the ACFT results. Moreover,
the dilaton always remains massless for ${\cal N}=1$ Minkowski vacua.
In the asymmetric Gepner models there exist an accompanying massless
NS-NS pseudo-scalar. If the $\tilde\phi$-field is not gauged this can be the
$\tilde\phi$-field itself or, as in the gauged case, it can in principle also be 
a linear combination of the NS-NS pseudo-scalars appearing
for the complex structure and complexified K\"ahler moduli.
By the flux also some of the complex structure moduli and 
complexified K\"ahler moduli are fixed. Since these 
appear in the NS-NS sector, the number of unconstrained ones 
should be compared with the number of scalars $N_0$ in the ACFT.

%%%%%%%%%%%%%%%%%%%%%%%%%%%%%%%%%%%%%%%%%%%%%%%
%%%%%%%%%%%%%%%%%%%%%%%%%%%%%%%%%%%%%%%%%%%%%%%

\subsubsection*{Remarks}

This  analysis has been performed in the framework of ${\cal N}=2$
GSUGRA, which, as argued before,  is not a priori  a fully established effective field
theory governing the dynamics of string theory on fluxed Calabi-Yau three-folds.
First, the question arises whether higher order corrections can induce
extra mass terms for the axions so that the bounds can be avoided.
Since the axions feature a perturbative continuous shift symmetry,
its potential is highly constrained. Like for axion-monodromy, perturbatively this shift
symmetry can only be broken by fluxes in a controlled way. As has been
argued in \cite{Kaloper:2008fb,Kaloper:2011jz,Bielleman:2015ina} all higher order corrections are expected to be corrections in
terms of the tree-level flux induced potential (instead of the axion field itself).
Secondly, there can be non-perturbative corrections for the axions.
Since axions arise in the R-R sector of string theory, these would be
D-brane instantons, i.e. non-perturbative effects in the string
coupling constant. These are not captured by the CFT, which only
incorporates world-sheet instantons, i.e. non-perturbative effects in
$\alpha'$. Following these arguments, the tree-level flux induced
potential for the R-R axions is expected to correctly capture the
dimension of the axionic moduli space.

However, there is a second more serious issue. In the asymmetric
Gepner models all massive modes are of order of the string mass.
Therefore, there is no  mass hierarchy among the Kaluza-Klein scale,
the string scale and the mass scale of the massive moduli. 
It is thus unclear whether a GSUGRA theory for the initially massless modes
can reliably describe the full dynamics of mass generation in string
theory. Of course gauge symmetry and shift-symmetry do protect masses
to a certain degree, but there can be subtle effects that for instance
generate masses for  the R-R axions via couplings to massive Kaluza-Klein
modes.

It is precisely one of the objectives  of this work to investigate to which
degree  minima of GSUGRA theories do provide  or at least hint at
true flux vacua of string theory. For that purpose let us now
consider concrete examples.

%%%%%%%%%%%%%%%%%%%%%%%%%%%%%%%%%%%%%%%%%%%%%%%
%%%%%%%%%%%%%%%%%%%%%%%%%%%%%%%%%%%%%%%%%%%%%%%
%%%%%%%%%%%%%%%%%%%%%%%%%%%%%%%%%%%%%%%%%%%%%%%
%%%%%%%%%%%%%%%%%%%%%%%%%%%%%%%%%%%%%%%%%%%%%%%
%%%%%%%%%%%%%%%%%%%%%%%%%%%%%%%%%%%%%%%%%%%%%%%
%%%%%%%%%%%%%%%%%%%%%%%%%%%%%%%%%%%%%%%%%%%%%%%

\section{ACFT -  GSUGRA correspondence} 
\label{sec_examples}

In this section we provide concrete examples of asymmetric Gepner
models and make an educated  proposal to  which  ${\cal N}=2$ GSUGRA 
these classical ${\cal N}=1$ Minkowski-type string vacua might
correspond to.
Of course, since the Gepner model is expected to lie deep inside the
K\"ahler moduli space and, in the ACFT, the backreaction
from all $\alpha'$-corrections has been taken into account, 
we cannot prove our conjecture but can 
only collect a number of indications for its correctness.

At the two  derivative level the effective action is given by ${\cal N}=2$
GSUGRA theory reviewed in the previous section. Let us recall 
the main features of partial supersymmetry breaking Minkowski vacua.
These should be considered as necessary conditions that  an
${\cal N}=1$ ACFT has to satisfy for admitting an interpretation as 
a fully adjusted minimum of an associated ${\cal N}=2$ GSUGRA theory.

\begin{enumerate}
\item{As we have seen, the dilaton and a  four-dimensional
    NS-NS pseudo-scalar $\varphi$ remain massless.}
\item{The constraint on the number of possible gaugings provided
the bound
\eq{
\label{bound_first}
                 h^{2,1}-h^{1,1}-\Delta\le      N_V\le h^{2,1}-1\,.
}
Since without gaugings/fluxes there are no charged fields under
the abelian R-R gauge symmetries, there seems to be no other
mechanism to make the $U(1)$s massive. Thus, we expect this bound to
strongly hold also in the ACFTs.
}
\item{Counting the number of massless axionic chiral multiplets led to
the two constraints
\eq{
\label{bound_second}
           N_V-N_{\rm ax}&\le h^{2,1}-h^{1,1}-\Delta\,,\\[0.2cm]
            N_V-2N_{\rm ax}&\ge h^{2,1}-2h^{1,1}-\Delta\,.
}
This was derived from the tree-level flux induced mass term for the
axions and was argued to hold even if higher derivative corrections
were taken into account in GSUGRA. However, there exist a  possible loop-hole:
Actually, the axionic Gepner models
feature no mass hierarchy between flux induced moduli masses and
the KK-scale. Therefore, axionic masses might be generated by
including KK modes. This would imply a smaller number of massless
axions relaxing the bound in the first relation  in \eqref{bound_second}.
}
\item{The number of massless scalars $N_0$ is the less constrained
    one, but at least it should satisfy
\eq{
 \label{bound_third}
             N_0\le h^{2,1}+h^{1,1}\,.
}
}         
\end{enumerate}

\noindent
Let us mention again that partial ${\cal N}=1$ Minkowski-type breaking requires
both magnetic and non-geometric gaugings/fluxes. Therefore,
if our conjecture is correct, we have identified asymmetric exactly solvable
classical string vacua containing in particular non-geometric fluxes.

\subsection{Procedure}

Let us describe how we proceed 
to identify a candidate GSUGRA model for an ACFT.
We start with a usual Gepner model with levels $(k_1,\ldots,k_5)$ 
that is known to correspond to a special point of a Calabi-Yau threefold
${\cal M}_{\rm Gep}$. This is usually a hypersurface in a weighted
projective space.
Then we extend  the modular invariant partition function by 
a simple current that does not commute with the GSO projection 
$J_{\rm   GSO}$ 
and sometimes also not with all of the additional simple currents $J_i$.
This gives the ACFT featuring 
a  number of massless vectors, R-R axions and scalars
$(N_V,N_{\rm ax};N_0)$ with $N_V>N_{\rm ax}$. The question now is 
whether one can find an ${\cal N}=2$ SUGRA defined on a Calabi-Yau manifold
${\cal M}_{\rm ACFT}$, whose gauging  admits an ${\cal N}=1$
Minkowski minimum that is related to  the ACFT model.

In order to identify a candidate for ${\cal M}_{\rm ACFT}$,
we take  a closer look at the massless vectors and try to understand
the combinatorics of these states. 
From this analysis we extract  an idea which weights the coordinates
and the constraints  of  ${\cal M}_{\rm ACFT}$ presumably have.
The difficulty is that one does not expect a one-to-one correspondence
between the massless vectors $N_V$ in the ACFT and
the massless vectors in the ungauged compactification on ${\cal
  M}_{\rm ACFT}$ leading to possible ambiguities. 
This is simply because due to additional gauging
some of the vectors of the ungauged theory become massive.
Once we isolated a candidate, we check whether the 
four conditions 1.-4. above are satisfied, i.e. in particular whether
the  relations
\eqref{bound_first}, \eqref{bound_second} and \eqref{bound_third}
hold. In order to avoid confusion, let us stress
that one gets ${\cal M}_{\rm ACFT}\ne {\cal M}_{\rm
  Gep}$. That additional simple currents usually change the underlying manifold can also be seen when using the D-type modular invariant by adding the simple current \eqref{Dsimplecurrent}. Our procedure can be  summarized as 
\vspace{0.2cm}
\begin{figure}[!ht]
 \centering
\includegraphics[width=0.7\textwidth]{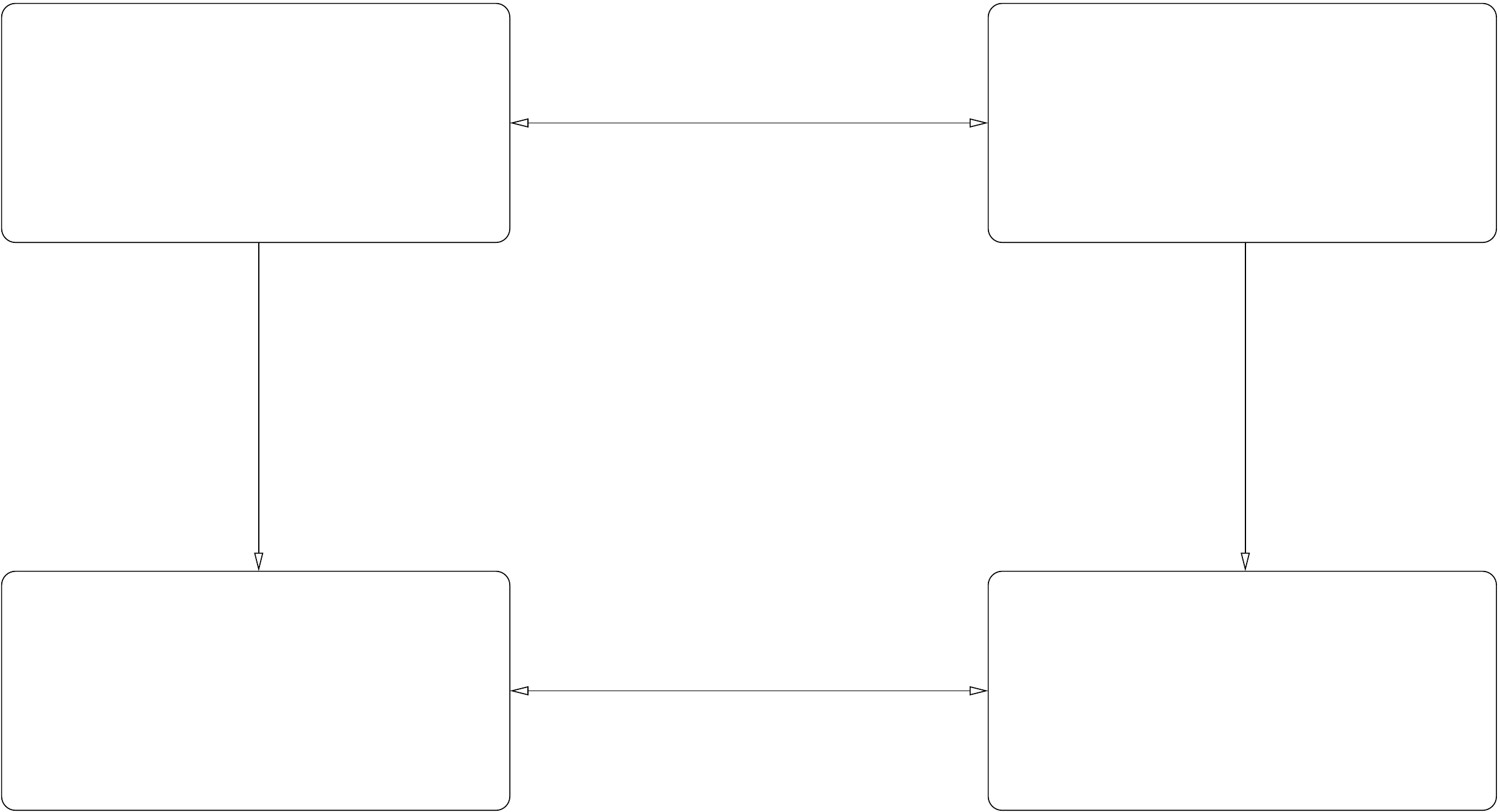}
 \begin{picture}(0,0)
      \put(-278,130){Gepner Model}
      \put(-87,130){SUGRA  ${\cal M}_{\rm Gep}$}
      \put(-262,21){ACFT}
      \put(-98,21){GSUGRA  $\!{\cal M}_{\rm ACFT}$}
      \put(-238,75){$J_{\rm ACFT}$}
    \end{picture}
\end{figure} 

\noindent
Now let us discuss a couple of examples that will clarify the just
described procedure.

\subsection{Odd level ACFT Models}
\label{sec_oddlev}

In this section we consider a special class of simple current
extended Gepner models $(k_1,\ldots,k_5)$.
We require that at least one of the levels, say $k_1$, is odd so that one
has the simple current
\eq{
\label{simple_asymgen}
           J_{\rm ACFT}=(0\;k_1+2\;1)(0\;0\;0)^4\,(c)\,.
}
Note that this simple current mixes the R and NS sector between the
different minimal models and is non-local with respect to  the $J_{i}$ in
\eqref{simplecurrents}.\footnote{Notice that this mixing does not
  necessarily break the left-moving $N=1$ world-sheet
  supersymmetry. For instance when replacing the $c$ by an $s$ in
  $J_{\rm ACFT}$ one finds models with $N=2$ target space SUSY which
  requires $N = (2,2)$ worldsheet supersymmetry. We think that what is
  happening here is the following: On each $N=2$ tensor factor one
   has a spectral flow operator characterized by a parameter $\eta_i$.
   The simple current $J_{\rm ACFT}$ (or with  $c$ and $s$ exchanged) act as a spectral flow
   operator with $\vec\eta=((k_1+2)/2,0,0,0,0;\pm 1/2)$. 
  If $G^\pm=\sum_i 1\otimes\ldots\otimes G^\pm_i\otimes\ldots\otimes
  1$ are the world-sheet supercurrents in the former Gepner model,
   then  $G^\pm_{\rm ACFT}=U_{\vec\eta}\, G^\pm\, U^\dagger_{\vec\eta}$
   become the left-moving world-sheet supercurrents in the  ACFT model.
   Since it is generated by a spectral flow, one has a full ``twisted''
   $N=2$ super Virasoro algebra.
}
Note that these are essentially the simple currents also
considered in the (0,2) heterotic Gepner models discussed in
\cite{Blumenhagen:1995tt,Blumenhagen:1995ew,Blumenhagen:1996vu}. Unlike there,  here we have the central charge $ c = (12,12)$ and therefore rather a type II string theory.

As the most simple example we first discuss
 the Gepner model with levels $(3,3,3,3,3)$ 
 extended by the simple current 
\eq{    
                J_{\rm ACFT}=(0\;5\;1)(0\;0\;0)^4\,(c)\,.
}
This breaks the left-moving supersymmetry and we obtain an ${\cal
  N}=1$ model with massless modes
\eq{
          1\times(\phi,\varphi)\quad + \quad     (N_V,N_{\rm ax}\,; N_0)=(80,0\,;74)\,.
}
To get an idea what this model might correspond to, we consider the
massless vectors in more detail. First, we note that the pure Gepner
model describes a special point in the moduli space of the quintic
${\cal M}_{\rm Gep}=\mathbb P_{1,1,1,1,1}[5]^{(101,1)}$. Since the simple current $J_{\rm ACFT}$ only
acts on the first factor, we expect that the other four coordinates
$x_i$ will still be present.
The $N_V=80$ massless modes are listed in table~\ref{table_B1}.
%%%%%%%%%%%%%%
%%%%%%%%%%%%%%
\begin{table}[ht]
\centering
\renewcommand{\arraystretch}{1.2}
\begin{tabular}{|c|c|c|}
  \hline
   state & polynom. rep.  & deg.\\
 \hline\hline
  $(0\;1\;1)(3\;4\;1)(2\;3\;1)(0\;1\;1)^2(s)$ & $x_i^3\, x_j^2$ &
  $12$\\
$(0\;1\;1)(3\;4\;1)(1\;2\;1)^2(0\;1\;1)(s)$ & $x_i^3\,
x_j\, x_k$ & $12$\\
$(0\;1\;1)(2\;3\;1)^2(1\; 2\; 1)(0\;1\;1)(s)$ & $x_i^2\,
x^2_j\, x_k$ & $12$\\
$(0\;1\;1)(2\;3\;1)(1\; 2\; 1)^3(s)$ & $x_i^2\,
x_j\, x_k\, x_l$ & $4$\\
\hline
$(1\;2\;1)(3\;0\;0)(0\;0\;0)^3(s)+$ &  & \\
$(2\;3\;1)(3\;4\;1)(0\;1\;1)^3(s)$ & $x_i^3\, y_m$ &
  $2\times 4=8$\\
$(1\;2\;1)(2\;0\;0)(1\;0\;0)(0\;0\;0)^2(s)+$ &  & \\
$(2\;3\;1)(3\;4\;1)(0\;1\;1)^3(s)$ & $x_i^2\, x_j \, y_m$ &
  $2\times 12=24$\\
$(1\;2\;1)(1\;0\;0)^3 (0\;0\;0)(s)+$ &  & \\
$(2\;3\;1)(1\;2\;1)^3(0\;1\;1)(s)$ & $x_i \, x_j\, x_k \, y_m$ &
  $2\times 4=8$\\
\hline
     \end{tabular} 
    \caption{\label{table_B1}  Combinatorics of the $N_V=80$ massless vectors.}
\end{table}
%%%%%%%%%%%%%%
%%%%%%%%%%%%%%
Note that in order the get the combinatorics right, besides the four
coordinates of weight one, $\{x_2,x_3,x_4,x_5\}$ we had to introduce two coordinates
$\{y_0,y_1\}$ of weight two.
Thus, all these 80 modes are given by the monomials of order $5$ divided
by an ideal
\eq{
                 {\cal P}_{5}\left( x_{i}\,,\, y_{j}
                 \right)/{\cal I}(x_i^4, y_j^2,  y_0 y_1 )\,.
}  
This observation motivates the following proposal for the underlying
(fluxed)  Calabi-Yau threefold
\eq{
        {\cal M}_{\rm ACFT}=  \mathbb P_{1,1,1,1,2,2}[5\;3] \,,
}
i.e. a complete intersection in a weighted projective
space. The degree three constraint has been introduced
to make it a Calabi-Yau three-fold. This of course introduces more
monomials than visible in the ACFT, but recall that due to the gauging
we cannot expect a one-to-one correspondence. We rather have to
satisfy the bounds \eqref{bound_first}, \eqref{bound_second} and \eqref{bound_third}.

The ambient space contains a $\mathbb Z_2$ singularity along
the curve ${\cal C}_2=\mathbb P^1$ that needs to be resolved.
This we do following the method described in appendix \ref{app_hodge}.
Using the intersection form  on the ambient space ${\cal A}$
\eq{
          I_{\cal A}={1\over 4} \eta^5
}
one computes 
\eq{
      \chi_F({\cal M}_{\rm ACFT})&=-\int_{{\cal A}} c_3(T_{\cal M})\,  15\eta^2=-{165}\,.
}
Thus, the Euler characteristic of the resolved CICY becomes
\eq{
   \chi({\cal M}_{\rm ACFT})&=\chi_F({\cal M}_{\rm ACFT}) - {\textstyle {1\over 2}} \chi( {\cal C}_2)
 +2 \chi( {\cal C}_2) =-162\,.
}
The resolution of the $\mathbb Z_2$ singularity introduces a single
additional K\"ahler modulus so that the Hodge numbers
of the CICY are  $(h^{2,1},h^{1,1})=(83,2)$. This agrees with
the result of the toric computation as listed in \cite{list_cicy} (see
also \cite{Kreuzer:2001fu,Klemm:2004km}).
Though we will discuss it in more detail in section \ref{sec_compare}, let us note that all four
necessary conditions 1.-4. are indeed satisfied.

This example can be generalized to an arbitrary Gepner model
with say the first level being an odd number $(2l-1,k_2,k_3,k_4,k_5)$.
This Gepner model corresponds to the Fermat-type  constraint
\eq{
              x_1^{2l+1}+x_2^{k_2+2}+x_3^{k_3+2}+x_4^{k_4+2}+x_5^{k_5+2}=0
}
in 
\eq{
{\cal M}_{\rm Gep}=
\mathbb P_{{d\over(2l+1)},{d\over k_2+2},{d\over k_3+2},{d\over k_4+2},{d\over k_5+2}}[d] \,,
}
with $d={\rm lcm}\{2l+1,k_2+2,k_3+2,k_4+2,k_5+2\}$.
Extension by the simple current \eqref{simple_asymgen} leads to an
asymmetric Gepner model, for which similar to
\cite{Blumenhagen:1995ew,Blumenhagen:1996vu} the vectors are given by the
combinatorics of polynomials of degree $d$ in the four coordinates $(x_2,x_3,x_4,x_5)$ of weights
$w_i=d/(k_i+2)$ and two new coordinates $(y_0,y_1)$ of weights 
$w_0=2d/(2l+1)$ and $w_1=ld/(2l+1)$.
Thus, we conjecture that this ACFT corresponds
to a Minkowski  vacuum of  the ${\cal N}=2$  GSUGRA 
on the CICY
\eq{
\label{cicymodelsa}
{\cal M}_{\rm ACFT}=\mathbb P_{{2d\over(2l+1)},{ld\over(2l+1)},{d\over k_2+2},{d\over
    k_3+2},{d\over k_4+2},{d\over k_5+2}}\left[d\ \; {\textstyle {d(l+1)\over (2l+1)}}\right]\,.
}
If we have only four tensor factors with say the last level being in
addition even, $k_4=2k$, then one can also choose for that factor
the D-type modular invariant by adding the simple current \eqref{Dsimplecurrent}. In this case the  Gepner model corresponds
 to the constraint
\eq{
              x_1^{2l+1}+x_2^{k_2+2}+x_3^{k_3+2}+x_4^{k+1}+x_4\, x_5^2=0
}
in 
\eq{
{\cal M}_{\rm Gep}=\mathbb P_{{d\over(2l+1)},{d\over k_2+2},{d\over k_3+2},{d\over k+1},{dk\over 2(k+1)}}[d]
}
with $d={\rm lcm}\{2l+1,k_2+2,k_3+2,k+1\}$.
For this class, the  asymmetric Gepner model should correspond
to a Minkowski  vacuum of  the ${\cal N}=2$  GSUGRA 
on the CICY
\eq{
\label{cicymodelsb}
{\cal M}_{\rm ACFT}=
\mathbb P_{{2d\over(2l+1)},{ld\over(2l+1)},{d\over k_2+2},{d\over
    k_3+2},{d\over k+1},{d k\over 2(k+1)}}\left[d\ \; {\textstyle {d(l+1)\over (2l+1)}}\right]\,.
}
In table \ref{table_2} we compare the massless spectrum of some of
these ACFTs  with the Hodge numbers of the CICYs, at least
for those cases where the CICY appeared in the list \cite{list_cicy}.
%%%%%%%%%%%%%%
%%%%%%%%%%%%%%
\begin{table}[ht]
\centering
\renewcommand{\arraystretch}{1.4}
\begin{tabular}{|c|c|c|}
  \hline
   Gepner & $(N_V,N_{\rm ax},N_0)$   &  CICY\\
 \hline\hline
$(3\,3\,3\,3\,3)$ & $(80,0,74)$ & $\mathbb P_{1,1,1,1,2,2}[4\;4]_{(83,2)}$
\\
$(5\,5\,5\,12_D)$ & $(86,2,80)$ & $\mathbb P_{1,1,1,2,3,3}[7\;4]_{(89,3)}$
\\
$(5\,5\,5\,12_A)$ & $(86,2,80)$ & $\mathbb P_{1,2,2,4,6,7}[14\;8]_{(88,4)}$
\\
$(7\,7\,7\,1\,1)$ & $(74,2,70)$ & $\mathbb P_{1,1,2,3,3,4}[9\;5]_{(75,6)}$ \\
\hline
     \end{tabular} 
    \caption{\label{table_2}  Examples: CFT-GSUGRA correspondence.}
\end{table}
%%%%%%%%%%%%%%
%%%%%%%%%%%%%%
In section \ref{sec_compare} we will compare the massless spectrum of
these asymmetric Gepner models with the expectation from GSUGRA
on the proposed Calabi-Yau three-folds. 

Let us emphasize that on the CFT side we have a plethora of consistent
models, but most of them cannot be directly related to a GSUGRA theory
on a large volume CICY. In fact, most of the spaces
\eqref{cicymodelsa} and \eqref{cicymodelsb} do not appear in the list of \cite{list_cicy}
so presumably do not yield transversally intersecting  CICYs.

%%%%%%%%%%%%%%
%%%%%%%%%%%%%%
\subsection{Level six ACFT Model}

In this subsection we consider the Gepner model with levels $(6_A,6_A,6_A,6_D)$ 
 extended by the simple current 
\eq{    
                J_{\rm ACFT}=(0\;4\;0)(0\;0\;0)^3\,(v)\,.
}
This simple current is non-local w.r.t. to the simple current $J_{\rm GSO}$ in \eqref{simplecurrents}, but being a pure NS-state $J_{\rm ACFT}$ is still local w.r.t. to all $J_{i}$. As a consequence the left-moving $N=1$ world-sheet supersymmetry is still manifest and one can find the usual $N=1$ supercurrents at the first massive level in the vacuum orbit. We obtain a model with ${\cal
  N}=1$ target-space supersymmetry with massless modes
\eq{
              (N_V,N_{\rm ax}\,; N_0)=(60,4\,;64) \,.
}
To get an idea what this model might correspond to, we consider the
massless vectors in more detail. First, we note that the pure Gepner
model describes a special point in the moduli space of the CY
${\cal M}_{\rm Gep}=\mathbb P_{1,1,1,2,3}[8]^{(106,2)}$. Since the simple current $J_{\rm ACFT}$ only
acts on the first factor, we expect that the four coordinates
$x_1,x_2,v,w$ of weights $(1,1,2,3)$ will still be present.
The $N_V=60$ massless modes are listed in table~\ref{table_B}.
%%%%%%%%%%%%%%
%%%%%%%%%%%%%%
\begin{table}[ht]
\centering
\renewcommand{\arraystretch}{1.2}
\begin{tabular}{|c|c|c|}
  \hline
   State & polynom. rep.  & deg.\\
 \hline\hline
  $(1\;2\;1)(6\;7\;1)(1\;2\;1)(0\;1\;1)(s)$ & $s_\alpha\, x_i^6\, x_j$ & $2\times
  2=4$\\
$(1\;2\;1)(5\;6\;1)(2\;3\;1)(0\;1\;1)(s)$ & $s_\alpha\,x_i^5\, x^2_j$ & $2\times
2=4$\\
$(1\;2\;1)(4\;5\;1)(3\;4\;1)(0\;1\;1)(s)$ & $s_\alpha\,x_i^4\, x^3_j$ & $2\times
2=4$\\
$(1\;2\;1)(3\;4\;1)(0\;1\;1)(0\;1\;1)(s)$ & $s_\alpha\,x_i^3\,z$ & $2\times 2=4$\\
$(1\;2\;1)(2\;3\;1)(1\;2\;1)(0\;1\;1)(s)$ & $s_\alpha\,x_i^2\, x_j\,z$ & $2\times 2=4$\\
\hline
 $(1\;2\;1)(5\;6\;1)(0\;1\;1)(2\;3\;1)(s)$ & $s_\alpha\,x_i^5\, v$ & $2\times
  2=4$\\
 $(1\;2\;1)(4\;5\;1)(1\;2\;1)(2\;3\;1)(s)$ & $s_\alpha\,x_i^4\, x_j\, v$ & $2\times
  2=4$\\
$(1\;2\;1)(3\;5\;1)(2\;2\;1)(2\;3\;1)(s)$ & $s_\alpha\,x_i^3\, x^2_j\, v$ & $2\times
  2=4$\\
$(1\;2\;1)(1\;2\;1)(0\;1\;1)(2\;3\;1)(s)$ & $s_\alpha\,x_i\, v\, z$ & $2\times
  2=4$\\
\hline
$(1\;2\;1)(4\;5\;1)(0\;1\;1)(3\;4\;1)(s)$ & $s_\alpha\,x_i^4\, w$ & $2\times
  2=4$\\
$(1\;2\;1)(3\;4\;1)(1\;2\;1)(3\;4\;1)(s)$ & $s_\alpha\,x_i^3\, x_j\, w$ & $2\times
  2=4$\\
$(1\;2\;1)(2\;3\;1)(2\;3\;1)(3\;4\;1)(s)$ & $s_\alpha\,x_1^2\, x_2^2\, w$ & $2\times
  1=2$\\
$(1\;2\;1)(0\;1\;1)(0\;1\;1)(3\;4\;1)(s)$ & $s_\alpha\,w\, z$ & $2\times
  1=2$\\
\hline
$(1\;2\;1)(3\;4\;1)(0\;1\;1)(4\;5\;1)(s)$ & $s_\alpha\,x_i^3\, v^2$ & $2\times
  2=4$\\
$(1\;2\;1)(2\;4\;1)(1\;2\;1)(4\;5\;1)(s)$ & $s_\alpha\,x_i^2\, x_j\, v^2$ & $2\times
  2=4$\\
\hline
$(1\;2\;1)(1\;1\;1)(0\;1\;1)(6\;7\;1)(s)$ & $s_\alpha\,x_i\, w^2(\sim v^3)$ & $2\times  2=4$\\
\hline
     \end{tabular} 
    \caption{\label{table_B}  Combinatorics of the $N_V=60$ massless vectors}
\end{table}
%%%%%%%%%%%%%%
%%%%%%%%%%%%%%
Note that all these ACFT states have a  twofold degeneracy which we can
capture by introducing a factor $s_{\alpha}$ with $\alpha=0,1$ into
the corresponding monomials. Moreover, to get the combinatorics right,
we introduced the new coordinate $z$ of weight four.
Therefore, all these 60 modes are given by the monomials of bi-order $[7,1]$ divided
by an ideal
\eq{
                 {\cal P}_{[7,1]}\left( x_{i\,[1,0]}\,,\, v_{[2,0]} \, ,\,
                     w_{[3,0]}\,,\, z_{[4,0]}\, ;\,
                     s_{\alpha\,[0,1]}\right)/{\cal I}(x_i^7, a v^3
                   +b w^2, vw, s_1 s_2)\,
}  
with $i,\alpha=0,1$.
This observation motivates the following proposal for the underlying
(fluxed)  Calabi-Yau threefold
\eq{
        {\cal M}_{\rm ACFT}=   \begin{matrix} \mathbb P_{1,1,2,3,4} \\ \mathbb P_{1,1}\hfill \end{matrix}\!
       \left[\begin{matrix}  7 & 4 \\ 1 & 1 \end{matrix}\right]\,,
}
i.e. a complete intersection in a product of weighted projective
spaces. 

Of course the first factor $\mathbb P_{1,1,2,3,4}$ features $\mathbb Z_2$,
$\mathbb Z_3$ and $\mathbb Z_4$ singularities. Let us discuss their
contribution to the Euler characteristic in more detail.
First, there exists a single point ${\cal  P}_4$ over which one has a
$\mathbb Z_4$ singularity.
Second, one has a $\mathbb Z_3$ singularity  over the curve
${\cal C}_3=\mathbb P^1$. 
Finally,  the CICY  has a $\mathbb Z_2$ singularity over a curve
 \eq{
         {\cal C}_2 =   \begin{matrix} \mathbb P_{1,2} \\ \mathbb P_{1,1}\hfill \end{matrix}\!
       \left[\begin{matrix}  2 \\  1 \end{matrix}\right]\,,
}
containing the former $\mathbb Z_4$ singularity. 
Using the methods described in  appendix~\ref{app_hodge},  we obtain for the Euler
characteristics of the singular loci
\eq{
            \chi({\cal  C}_3)&=2\,,\qquad \chi({\cal P}_4)=1 \,, \\[0.1cm]
            \chi( {\cal C}_2/{\cal P}_4)&=\chi_F({\cal C}_2)
            -{\textstyle {1\over 2}}\chi({\cal P}_4)=1\,.
}
Using the intersection form  on the ambient space ${\cal A}$
\eq{
          I_{\cal A}={1\over 24} \eta_1^4\, \eta_2
}
one computes 
\eq{
      \chi_F({\cal M}_{\rm ACFT})&=-\int_{{\cal A}} c_3(T_{\cal M})\, (7\eta_1+\eta_2)(4\eta_1+\eta_2)=-{1471\over 12}\,.
}
Thus, the Euler characteristic of the resolved CICY becomes
\eq{
   \chi({\cal M}_{\rm ACFT})&=\chi_F({\cal M}_{\rm ACFT}) - {\textstyle {1\over 2}} \chi( {\cal C}_2/{\cal
     P}_4) - {\textstyle {1\over 3}} \chi( {\cal C}_3) -
   {\textstyle {1\over 4}} \chi( {\cal P}_4)\\[0.1cm]
   &\hspace{2.8cm} + 2 \chi( {\cal C}_2/{\cal   P}_4) +3 \chi( {\cal C}_3) +4 \chi(
   {\cal P}_4)\\
  &=-112\,.
}
Moreover, besides the two toric K\"ahler classes, the resolution
introduces the following number of extra K\"ahler classes
\eq{    
      \mathbb Z_2: h^{1,1}=1\,,\qquad \mathbb Z_3: h^{1,1}=2\,,\qquad 
    \mathbb Z_4: h^{1,1}=1 \,,
}
so that the Hodge numbers of the resolved CICY are $(h^{2,1},h^{1,1})=(62,6)$.

\subsection{Level ten ACFT model}

Finally let us briefly discuss  the Gepner model $(10, 10, 4, 4)$ with only
$A$-type modular invariants that corresponds
to the Calabi-Yau ${\cal M}_{\rm Gep}=\mathbb{P}_{1,1,2,2,6}[12]$.
We extend the partition function again by the simple current 
\eq{    
                J_{\rm ACFT}=(0\;4\;0)(0\;0\;0)^3\,(v)\,,
}
yielding an ${\cal N}=1$ Minkowski vacuum with massless modes
\eq{
              (N_V,N_{\rm ax}\,; N_0)=(59,5\,;68) \,.
}
In terms of the three unaffected coordinates $\{x_2,x_3,x_4\}$ of
weights $\{1,2,2\}$ all the massless states follow the combinatorics
shown in Table \ref{table_CC}.

%%%%%%%%%%%%%%
%%%%%%%%%%%%%%
\begin{table}[ht]
\centering
\renewcommand{\arraystretch}{1.2}
\begin{tabular}{|c|c|}
  \hline
   polynom. rep.  & deg.\\
 \hline\hline
  $p_{11}(x)$ & $18$\\
  $p_{10}(x)$ & $19$\\
   $p_{7}(x)$ & $10+2$\\
   $p_{4}(x)$ & $6$\\
 $p_{3}(x)$ & $3$\\
\hline
     \end{tabular} 
    \caption{\label{table_CC}  Combinatorics of the $N_V=59$ massless
      vectors in terms of polynomials $p_n(x_2,x_3,x_4)/{\cal I}(x_2^{11},x_3^{5},x_4^5)$.}
\end{table}
%%%%%%%%%%%%%%
%%%%%%%%%%%%%%
Furthermore one sees that the states fall into a twisted and an untwisted sector. Arranging the degrees according to these sectors into two sets $\{11,7,3\}$ and $\{10,7,4\}$, 
one might be tempted to introduce two extra coordinates of weights
three and four. Taking also the maximally appearing  degrees $11$ and
$10$
into account, we conjecture that the underlying Calabi-Yau threefold
could be
\eq{
             {\cal M}_{\rm ACFT}=\mathbb P_{1,2,2,3,4,9}[11\ 10]\,.
}
Of course some of the possible monomials appearing in  ${\cal M}_{\rm
  ACFT}$ are missing on the ACFT side, but that is expected due to the gauging.
Employing the methods from appendix \ref{app_hodge} we 
can derive the Hodge numbers $(h^{2,1},h^{1,1})=(66,8)$.
At least these numbers lie in the right ballpark.
However, this example shows that it is not straightforward to identify 
a candidate GSUGRA model. It always involves a bit of guess-work and
intuition.

\subsection{Check of GSUGRA constraints}
\label{sec_compare}

In this section we check whether the massless spectra for the
asymmetric Gepner models are consistent
with the necessary constraints 1.-4. from the effective GSUGRA
analysis. To explicitly check whether there exist concrete choices for
the fluxes that admit these GSUGRA vacua is a difficult though interesting question.

First of all, all presented examples of ACFTs  do contain
a universal massless chiral multiplet from the vacuum orbit. 
This is the candidate for hosting the dilaton and a NS-NS 
pseudo-scalar. In GSUGRA we have seen that the latter could
be the $\tilde\phi$-field itself or a linear combination of the many
NS-NS pseudo-scalars residing in the complex structure
and complexified K\"ahler moduli. 
Second, for GSUGRA we have derived the bounds \eqref{bound_first},
and \eqref{bound_second}
on the number of massless vector and chiral RR-axion multiplets. 
The mild bound \eqref{bound_third} on the number of scalars 
is satisfied for all examples.
In table \ref{table_4} we compare the ACFT data with the GSUGRA bounds.
%%%%%%%%%%%%%%
%%%%%%%%%%%%%%
\begin{table}[p] 
\centering
\renewcommand{\arraystretch}{1.3}
\begin{tabular}{|c|c|c|c|}
  \hline
   Gepner & $(N_V,N_{\rm ax},N_0)$   &  $(h^{2,1},h^{1,1})$ & constraints \\
 \hline\hline
$(3\,3\,3\,3\,3)$ & $(80,0,74)$ & $(83,2)$ & {\small $N_V-N_{\rm ax}\le 81-\Delta$} \\
& & &   {\small $N_V-2N_{\rm ax}\ge 79-\Delta$} \\
& & &   {\small $ 81-\Delta \le N_V\le 82$} \\ \hline
$(5\,5\,5\,12_D)$ & $(86,2,80)$ & $(89,3)$ & {\small $N_V-N_{\rm
    ax}\le 86-\Delta$} \\
& & &   {\small $\dashuline{N_V-2N_{\rm ax}\ge 83-\Delta}$} \\
& & &   {\small $ 86-\Delta \le N_V\le 88$} \\ \hline
$(5\,5\,5\,12_A)$ & $(86,2,80)$ & $(88,4)$ & {\small $N_V-N_{\rm
    ax}\le 84-\Delta$} \\
& & &   {\small $N_V-2N_{\rm ax}\ge 80-\Delta$} \\
& & &   {\small $ 84-\Delta \le N_V\le 87$} \\ \hline
$(7\,7\,7\,1\,1)$ & $(74,2,70)$ & $(75,6)$  & {\small $\underline{N_V-N_{\rm
    ax}\le 69-\Delta}$} \\
& & &   {\small $N_V-2N_{\rm ax}\ge 63-\Delta$} \\
& & &   {\small $ 69-\Delta \le N_V\le 74$} \\ \hline
$(6\,6\,6\,6_D)$ & $(60,4,64)$ & $(62,6)$  &  {\small $N_V-N_{\rm
    ax}\le 56-\Delta$} \\
& & &   {\small $N_V-2N_{\rm ax}\ge 50-\Delta$} \\
& & &   {\small $ 56-\Delta \le N_V\le 61$} \\\hline
$(10\,10\,4\,4)$ & $(59,5,68)$ & $(66,8)$  &  {\small $N_V-N_{\rm
    ax}\le 58-\Delta$} \\
& & &   {\small $\dashuline{N_V-2N_{\rm ax}\ge 50-\Delta}$} \\
& & &   {\small $ 58-\Delta \le N_V\le 65$}\\
\hline
     \end{tabular} 
    \caption{\label{table_4}  Check: ACFT-GSUGRA correspondence. 
The  underlined condition is not satisfied while the dashed conditions are satisfied if only one of the RR-axions $\zeta_{1,2}$ remains massless after being fixed by \eqref{RRfixing}.}
\end{table}
%%%%%%%%%%%%%%
%%%%%%%%%%%%%%
One realizes that the bounds are fairly strong, not leaving much freedom
for  the number of massless vectors and R-R axions. For instance
for the asymmetric $(3)^5$ Gepner model the GSUGRA constraints only
admit the six possible spectra given by $(N_V,N_{\rm ax})\in\{(80,0),(80,1),(81,0),(81,1),(82,1),(82,2)\}$.

For the number of vector multiplets the bounds are always satisfied.
Recall that this is certainly the mostly protected sector.
As indicated in Table \ref{table_4}, there exists one case where
the GSUGRA conditions for the RR-axions are  not satisfied.  
For the asymmetric $(1^2\, 7^3)$ Gepner model, 
the GSUGRA predicts too many massless R-R axions. Therefore, this 
models seems to need some dynamics that is not captured by ${\cal N}=2$
GSUGRA.

Apart from that we consider the correspondence
between the massless spectra of ${\cal N}=1$ asymmetric Gepner models
and partially broken ${\cal N}=2$ GSUGRA very encouraging and would
like to conjecture that  the ACFTs do really describe the fully
backreacted solutions, that are indicated by Minkowski vacua
of an GSUGRA approximation. 

Again, we are not claiming that the latter
gives a completely established Wilsonian effective
description. Instead, as argued in
\cite{Kaloper:2008fb,Kaloper:2011jz,Bielleman:2015ina}, large parts of
the full dynamics are dictated by the tree level potential and might
be  protected enough such that there are indeed solutions that
survive in the full string theory after adjusting themselves.

\clearpage

\section{Conclusions}
\label{sec_concl}

In this paper we have collected some evidence that a certain class
of asymmetric Gepner models can be identified with fully
backreacted vacua, that are indicated by partially broken
${\cal N}=1$ Minkowski minima of corresponding ${\cal N}=2$
GSUGRA theories. Our work goes 
beyond the former attempts \cite{Schellekens:1989wx,Israel:2013wwa,Israel:2015efa}
in that we took a closer look at the massless states in the ACFTs
and came up with concrete proposals for the CICYs underlying
the GSUGRA theories. 
It is almost inevitable that there exists some ambiguity in the choice
of the underlying CY manifold,
as due to the flux some of the axions and scalars have already become
massive. Since for our examples there seems to be only few moduli
missing, the number of fluxes turned on is expected to be rather
small. We derived a number of constraints for the massless spectra for 
${\cal N}=1$ minima of GSUGRA that were almost all satisfied
by the candidates for the ACFT-GSUGRA correspondence.
Moreover,  this picture  fits perfectly with the expectation that non-geometric fluxes are related to
asymmetric CFTs.

Clearly, we were just collecting arguments but could not give a
complete proof of our conjecture.
It would be desirable to be more concrete about precisely which fluxes have been turned
on,  but that requires the knowledge of the period matrices of complex structure
and K\"ahler moduli in the vicinity of a small radius Gepner point.
For the CICYs appearing in our list, this is not known.

The class of ACFT that we were considering is huge and only very few
models
could be identified with large volume geometries. In general we expect
that these ACFTs only exist in the stringy regime not admitting
any geometric interpretation.

\vskip2em
\noindent
\emph{Acknowledgments:} We are grateful to S. Greiner, T. W. Grimm, D. Junghans and
E. Malek  for helpful discussions.

%%%%%%%%%%%%%%%%%%%%%%%%%%%%%%%%%%%%%%%%%%%%%%%
%%%%%%%%%%%%%%%%%%%%%%%%%%%%%%%%%%%%%%%%%%%%%%%
%%%%%%%%%%%%%%%%%%%%%%%%%%%%%%%%%%%%%%%%%%%%%%%
%%%%%%%%%%%%%%%%%%%%%%%%%%%%%%%%%%%%%%%%%%%%%%%

\clearpage
\appendix
\section{Hodge numbers for CICY}
\label{app_hodge}

In this appendix we review the employed  technique to determine
the Hodge numbers of  CICYs in products  of weighted 
projective spaces \cite{Fuchs:1989pt,Fuchs:1989yv}. The latter are generically singular so that 
one has to resolve them.

Say we want to compute the Hodge numbers of $\mathbb
P_{w_1,\ldots,w_6}[d_1\; d_2]$.
The intersection form in the in general singular ambient space
is
\eq{
              I_{\cal A}={1\over \prod_i w_i} \eta^5\,.
}
Whenever some of the coordinates have a common divisor $N$,
one has a $\mathbb Z_N$ singularity over the locus where
the remaining coordinates vanish. As long as the intersection
of these loci with the two hypersurface constraints leads to  
singular points and curves ${\cal D}$, the CICY can be resolved in a Ricci-flat
manner.
The Euler characteristic of the resolution can be computed via
\eq{
    \chi(M)=\chi_F(M_{\rm sing})-{1\over N}\chi({\cal D})+N\chi({\cal D})
}
where the rational number $\chi_F(M_{\rm sing})$ can be computed via
\eq{
      \chi_F(M_{\rm sing})&=-\int_{{\cal A}} c_3(T_M)\, (d_1 d_2\, \eta^2)\,.
}
The third  Chern class can be read off from the total Chern class 
\eq{
     c(T_M)={ (1+d_1 \eta)(1+d_2\eta)\over \prod_i (1+w_i \eta)}\Big\vert_{\eta^5} \,.
}
It often happens that the various singularities do intersect.
In this case the above formulas have to be iterated such that
each singularity is only counted ones. How this works, is demonstrated
for the examples explicitly discussed the main text of this paper.

The number of K\"ahler classes can be computed in the
following way. Besides the canonical $(1,1)$ forms
inherited from the ambient space, from the resolution
of singular curves and points one gets:
\begin{itemize}
\item{The resolution of a singular curve of order $N$ introduces
     $(N-1)$ additional $(1,1)$-forms.}
\item{The resolution of a singular point of order $N$ introduces
     ${1\over 2}(N-1)$ additional $(1,1)$-forms.}
\item{If on top of a singular curve of order $N$ there are singular
    points of order $N \cdot M$ for each such point one deducts 
    ${1\over 2}(N-1)$ $(1,1)$-forms.}
\end{itemize}

%%%%%%%%%%%%%%%%%%%%%%%%%%%%%%%%%%%%%%%%%%%%%%%
%%%%%%%%%%%%%%%%%%%%%%%%%%%%%%%%%%%%%%%%%%%%%%%
%%%%%%%%%%%%%%%%%%%%%%%%%%%%%%%%%%%%%%%%%%%%%%%
%%%%%%%%%%%%%%%%%%%%%%%%%%%%%%%%%%%%%%%%%%%%%%%

\clearpage
\bibliography{references}  
\bibliographystyle{utphys}

%%%%%%%%%%%%%%%%%%%%%%%%%%%%%%%%%%%%%%%%%%%%%%%
%%%%%%%%%%%%%%%%%%%%%%%%%%%%%%%%%%%%%%%%%%%%%%%
%%%%%%%%%%%%%%%%%%%%%%%%%%%%%%%%%%%%%%%%%%%%%%%
%%%%%%%%%%%%%%%%%%%%%%%%%%%%%%%%%%%%%%%%%%%%%%%

\end{document}